\documentclass{
JHEP3}

\usepackage{latexsym}
\usepackage{amssymb}
\usepackage{array}

\def\0{\over } \def\1{\vec } \def\2{{\textstyle{1\over2}}} 
\def\3#1{{\bf{#1}}}
\def\4{{\textstyle{1\over4}}}
\def\5{\bar } 
\def\6{\partial }
\def\7#1{{#1}\!\llap{/}\,}
\def\8#1{{\textstyle{#1}}} 
\def\9{\underline }
\def\.{\cdot }
\def\^#1{\widehat{#1}}
\def\olrD{\,\hbox{\raisebox{2ex}{$\scriptstyle \leftrightarrow$}} \llap{$D$}}
\def\olrp{\hbox{\raisebox{2ex}{$\scriptstyle \leftrightarrow$}} \llap{$\partial$}}
\def\olp{\hbox{\raisebox{2ex}{$\scriptstyle \leftarrow$}} \llap{$\partial$}}

 \let\G=\Gamma  

\def\intx{\int\!d^3x\, }

\def\CL{{\cal L}}

\def\J{{\cal J}}

\newcommand{\pv}{\phi_{\mathrm V}}

\def\({\left(} \def\){\right)} \def\<{\langle } \def\>{\rangle }
\def\[{\left[} \def\]{\right]}  
 
\def\pmbf#1{\setbox0=\hbox{${#1}$}
        \kern-.025em\copy0\kern-\wd0
        \kern.05em\copy0\kern-\wd0
        \kern-.025em\raise.0433em\box0 }

\def\be{\begin{equation}}
\def\ee{\end{equation}}

\newcommand{\bel}[1]{\begin{equation}\label{#1}}
\def\bea{\begin{eqnarray}}
\newcommand{\beal}[1]{\begin{eqnarray}\label{#1}}
\def\eea{\end{eqnarray}}
\def\nn{\nonumber\\ }

\newcommand{\unit}{\mathbf1}
\newcommand{\intsum}{\sum\!\!\!\!\!\!\!\int}  

\def\tr{\,{\rm tr}\,}

\newcommand{\slasha}[1]{#1\! \! \! / }

\newcommand{\hal}{{\textstyle\frac{1}{2}}}

\newcommand{\bg}[1]{{{#1}}}

\newcommand{\mpar}[1]{}

\preprint{\today\\TUW-05-22\\YITP-SB-05-34\\ITP-UH-34/05}

\title{QUANTUM MASS AND CENTRAL CHARGE\\
OF SUPERSYMMETRIC MONOPOLES\\
Anomalies, current renormalization, and surface terms}

\author{Anton Rebhan\\
Institut f\"ur Theoretische Physik, Technische Universit\"at Wien,\\
  Wiedner Hauptstr. 8--10, A-1040 Vienna, Austria}

\author{Peter van Nieuwenhuizen\\
C.N.Yang Institute for Theoretical Physics,
  SUNY at Stony Brook,\\ Stony Brook, NY 11794-3840, USA } 

\author{Robert Wimmer\\
Institut f\"ur Theoretische Physik, Universit\"at Hannover,\\
Appelstr.~2, D-30167 Hanover, Germany}

\abstract{We calculate the one-loop quantum corrections to the mass and central
charge of $N=2$ and $N=4$ supersymmetric monopoles
in 3+1 dimensions.
The corrections to the $N=2$ central charge are finite and
due to an anomaly in the conformal central charge current,
but they cancel for the $N=4$ monopole.
For the quantum corrections to the mass we start with
the integral over the expectation value of the
Hamiltonian density, which we show to consist of a bulk contribution
which is given by the familiar sum over zero-point energies,
as well as surface terms which contribute nontrivially in the
monopole sector.
The bulk contribution is evaluated through index theorems
and found to be nonvanishing only in the $N=2$ case.
The contributions from the
surface terms
in the Hamiltonian are cancelled by infinite
composite operator counterterms
in the $N=4$ case,
forming a multiplet of improvement terms.
These counterterms are also needed for the renormalization of the central
charge. However, in the $N=2$ case they
cancel, and both the improved and the unimproved
current multiplet are finite.}

\begin{document}

\section{Introduction}

The existence of
supersymmetric (susy) monopoles \cite{D'Adda:1978ur,D'Adda:1978mu} 
which saturate the Bogomolnyi bound \cite{Bogomolny:1976de}
also at the quantum level \cite{Witten:1978mh}
plays an important role in the successes of nonperturbative studies
of super Yang-Mills theories through dualities \cite{Seiberg:1994rs,Seiberg:1994aj,Alvarez-Gaume:1997mv,Tong:2005un}. 

On the other hand, a direct calculation of quantum corrections
to the mass and central charge of susy solitons has proved to
be fraught with difficulties and surprises.
While in the earliest literature it was assumed that supersymmetry
would lead to a complete cancellation of quantum corrections
to both \cite{D'Adda:1978ur,D'Adda:1978mu,Rouhani:1981id}, 
it was quickly realized
that the bosonic and fermionic quantum fluctuations
do not only not cancel, but have to match the infinities
in standard coupling and field renormalizations \cite{Schonfeld:1979hg,Kaul:1983yt,Chatterjee:1984xh,Yamagishi:1984zv,Uchiyama:1984kb,Imbimbo:1984nq}.
However, even in the simplest case of the 1+1 dimensional
minimally supersymmetric kink, there was until the end of the 1990's
an unresolved discrepancy in the literature as to the precise
value of one-loop contributions once the renormalization scheme has
duly been fixed. As pointed out in \cite{Rebhan:1997iv}, 
most workers had used regularization
methods which when used naively give inconsistent results
already for the exactly solvable sine-Gordon kink.
In the susy case, there is moreover an extra complication
in that the traditionally employed periodic boundary conditions
lead to a contamination of the results by energy located at
the boundary of the quantization volume, and the issue
of the correct quantum mass of the susy kink was finally
settled in Ref.~\cite{Nastase:1998sy} by the use of topological boundary
conditions, which avoid this contamination.\footnote{Ref.~\cite{Nastase:1998sy}
used ``derivative regularization'' to make this work. In 
mode regularization it turns out that one has to average over
sets of boundary conditions to cancel both localized
boundary energy and delocalized momentum 
\cite{Goldhaber:2000ab,Goldhaber:2002mx}.}
This singled out as correct the earlier result
of Ref.~\cite{Schonfeld:1979hg,Casahorran:1989vd}
and refuted the null results of 
Refs.~\cite{Kaul:1983yt,Chatterjee:1984xh,Yamagishi:1984zv,Uchiyama:1984kb}.
However it led to a new problem because it seemed 
that the central charge did not appear to receive corresponding quantum
corrections \cite{Imbimbo:1984nq}, which would
imply a violation of the Bogomolnyi bound.
In Ref.~\cite{Nastase:1998sy}
it was conjectured that a new kind
of anomaly was responsible, and in Ref.~\cite{Shifman:1998zy}
Shifman, Vainstein, and Voloshin 
subsequently demonstrated that supersymmetry requires an anomalous
contribution to the central charge current.\footnote{Refs.\
\cite{Graham:1998qq,Casahorran:1989nr}, who had obtained
the correct value for the quantum mass also claimed
a nontrivial quantum correction
to the central charge apparently without the need of the anomalous term
proposed in Ref.~\cite{Shifman:1998zy}. However,
as shown in Ref.~\cite{Rebhan:2002yw}, this was achieved by formal
arguments handling ill-defined since unregularized quantities.}
The latter appears
in the same multiplet as the trace and conformal-susy anomalies,
and ensures BPS saturation even in the $N=1$ susy kink,
where initially standard multiplet shortening arguments seemed not
to be applicable.\footnote{That multiplet shortening also
occurs in the $N=1$ susy kink was eventually clarified
in Ref.\ \cite{Losev:2000mm}.}

In Ref.~\cite{Rebhan:2002uk} we have developed a version
of dimensional regularization which can be used
for solitons (and instantons).
The soliton is embedded in a higher-dimensional space
by adding extra trivial dimensions \cite{Luscher:1982wf,Parnachev:2000fz}
and choosing a model which is supersymmetric in the bigger space
and which reproduces the original model by dimensional reduction.
This is thus a combination of standard 't Hooft-Veltman
dimensional regularization \cite{'tHooft:1972fi} which goes up in dimensions,
and susy-preserving dimensional reduction
\cite{Siegel:1979wq,Capper:1980ns}, which goes down.
In Ref.~\cite{Rebhan:2002yw}
we demonstrated how the {anomalous contribution to} the central charge
of the susy kink can be obtained as a remnant of parity violation
in the odd-dimensional model used for embedding the susy kink,
and recently we showed that the same kind of anomalous contribution
arises in the more prominent case of the 3+1-dimensional monopole
of $N=2$ super-Yang-Mills theory in the Higgs\footnote{Anomalous
contributions to the central charge appear also in the newly
discovered ``confined monopoles'' pertaining to the
Coulomb phase \cite{Shifman:2003uh,Shifman:2004dr}, 
which turn out to be related to central charge
anomalies of 1+1-dimensional $N=2$ sigma models with twisted mass
\cite{Shifman:2006bs}.}
phase \cite{Rebhan:2004vn}.
This previously overlooked \cite{Kaul:1984bp,Imbimbo:1985mt}
finite contribution turns out to be in fact essential 
for consistency of these direct calculations
with the $N=2$ low-energy effective action
of Seiberg and Witten 
\cite{Seiberg:1994rs,Seiberg:1994aj,Alvarez-Gaume:1997mv}.
(We
have also found previously overlooked finite contributions to
both mass and central charge of the $N=2$ vortex in 2+1 dimensions
\cite{Vassilevich:2003xk,Rebhan:2003bu},
which are however not associated with conformal anomalies
but are rather standard renormalization effects.)

In the case of the $N=4$ monopole, it turns out that
the quantum corrections to the mass and central charge are
anomaly-free in accordance with the finiteness of the
model and the vanishing of
the trace and conformal-susy anomalies. However, a direct
calculation of these corrections leads to the surprising
appearance of composite-operator counterterms that
were absent in the $N=2$ case. This issue was recently
clarified in Ref.~\cite{Rebhan:2005yi}, where it was found
that in the 3+1-dimensional cases there is generally
a need for composite-operator counterterms forming
a multiplet of improvement terms, except when $N=2$.

In this paper we give a unified treatment of the $N=2$ and
$N=4$ monopoles and present direct and complete one-loop calculations
of anomalous and nonanomalous contributions to their
quantum mass and central charge.
This requires a careful calculation of both bulk contributions
and surface terms. The anomalous contributions arise from the
bulk, which in the case of the mass are given by
sums over zero-point energies. In the central charge, on the other hand,
such bulk contributions appear through momentum operators
in the extra dimensions introduced by our method of dimensional
regularization. Both types of bulk contributions
can be evaluated through the use of index theorems, which express
nontrivial differences for the spectral densities
of bosonic and fermionic contributions
in terms of the axial anomaly \cite{Callias:1977kg,Weinberg:1979ma,Imbimbo:1985mt}.
Surface terms, which at the classical level yield both the central charge 
and the mass, can also produce quantum corrections to 
the mass and the central charge
in the monopole sector, in stark contrast to the situation
in lower-dimensional models (which do not involve massless fields).
The traditional classical value of the monopole mass 
and central charge is obtained
when (and only when) unimproved currents are used, which are
singled out when the
$N=2$ and $N=4$ super-Yang-Mills theories
are derived from dimensional reduction of
the $N=1$ super-Yang-Mills theory in 5+1 and 9+1 dimensions,
respectively. Concerning quantum corrections due to the surface terms,
the $N=2$ monopole is special in that both unimproved and
improved currents are finite, but in the $N=4$ monopole
the unimproved current multiplet requires additive infinite composite-operator
renormalization through a multiplet of improvement terms.

After setting up the models in question in Section 2 and
presenting their solitonic solutions, the susy algebra
and the associated currents, we discuss quantization, regularization and
renormalization of these models in Section 3.
In Section 4 we start our direct calculation of the
monopole mass by taking the integral over the expectation value
of the energy density, and separating bulk and surface contributions
so that the former correspond to the familiar sums over
zero-point energies. In Section 5 we derive and discuss the index theorems
for evaluating the latter
\cite{Weinberg:1979ma,Weinberg:1981eu}, and apply them to all
lower-dimensional models as well as the monopole and instantons.
In Section 6 we evaluate the quantum corrections to the mass
arising from the surface contributions. These
turn out to cancel in the $N=2$ case, but require infinite
renormalization in the otherwise finite $N=4$ theory.
By combining the anomalous and nonanomalous contributions
we obtain the final result for the
$N=2$ and $N=4$ quantum mass. In Section 7, we perform the analogous direct
evaluation of the expectation value of the central charges.
Here the anomalous contribution of the $N=2$
model is identified
as the remnant of parity violation in the 4+1-dimensional
theory that one can use for trivial embedding of a monopole
as a string-like object without
violating supersymmetry. 
Section 8 
contains our conclusions
and some assorted comments on technical issues encountered
in calculations of the quantum corrections to
mass and central charge of susy solitons.

\section{The N=2 and N=4 susy monopole models
and their supercurrents}

In this section we discuss the $N=2$ and $N=4$ super Yang-Mills
models with monopoles, their BPS equations, susy algebra, improvement
terms and the susy variation of the latter. We begin with the
$N=2$ case, and then present the $N=4$ case.

\subsection{N=2}

The $N=2$ super Yang-Mills theory in 3+1 dimensions can be obtained
by dimensional reduction from the simplest ($N=1$) super Yang-Mills
theory in 5+1 dimensions \cite{Brink:1977bc}. 
Instead of using two complex chiral spinors
with a symplectic Majorana condition in order
to exhibit the R symmetry group U(2) of the $N=2$ susy algebra
in 3+1 dimensions \cite{Sohnius:1985qm}, we use the simpler
formulation with one complex chiral Dirac field $\lambda$.
The Lagrangian in 5+1 dimensions reads
\begin{equation}
  \label{eq:L6d}
  \mathcal L
=- \4 F_{MN}^a F^{aMN} - \bar{\lambda}^a\Gamma^M (D_M\lambda)^a,
\end{equation}
where the indices $M,N$ take the values $0,1,2,3,5,6,$
and the metric has signature $(-,+,\ldots,\break +)$.
The Dirac matrices 
are $8\times8$ matrices, and only $\Gamma^0$ is
anti-hermitian. Defining
$\bar\lambda^a=(\lambda^a)^\dagger i \Gamma^0$,
the action is hermitian without
an extra factor of $i$ in front of the fermionic part
(up to fermionic surface terms which do not contribute,
not even in the solitonic sector).
The covariant derivative $(D_M\lambda)^a$
is defined by $\6_M\lambda^a+g f^{abc}A_M^b \lambda^c$.
We use SU(2) as gauge group with
$f^{abc}=\epsilon^{abc}$. 
Since $\lambda^a$ is in the adjoint representation,
there is no difference between $\5\lambda^a$ and $\5\lambda_a$.

The chirality condition can be written as
\begin{equation}
  \label{eq:wcon}
  (1-\Gamma_7)\lambda=0\quad \textrm{with} \quad 
  \Gamma_7=\Gamma_0\Gamma_1\Gamma_2\Gamma_3\Gamma_5\Gamma_6 
\end{equation}
and $\Gamma_7^2=1$.
To carry out the dimensional reduction we write
$A_M=(A_\mu,P,S)$ and choose the following representation of
gamma matrices
\begin{eqnarray}
  \label{eq:6dgam}
  \Gamma_\mu=\gamma_\mu\otimes\sigma_1\ &,&\ \mu=0,1,2,3,\nonumber\\
  \Gamma_5=\gamma_5\otimes\sigma_1\ &,&\ \Gamma_6=\unit\otimes\sigma_2
\end{eqnarray}
where $\gamma_5=\gamma^1\gamma^2\gamma^3i\gamma^0$ and $\gamma_5^2=\mathbf 1$.
In this representation
the Weyl condition (\ref{eq:wcon}) becomes 
$\lambda={\psi\choose 0}$, 
with a complex four-component spinor $\psi$.
The six-dimensional charge conjugation matrix 
satisfying $C_{6} \Gamma_M C_{6}^{-1}=-\Gamma_M^T$
is $C_{6}=C\gamma_5 \otimes
\sigma_2$ where $C$ is the four-dimensional 
charge conjugation matrix satisfying 
 $C\gamma_\mu C^{-1}=-\gamma_\mu^T$.
Since $C^T=-C$ and $(C\gamma_5)^T=-C\gamma_5$ one has $C_{6}^T=+C_{6}$.

The
(3+1)-dimensional Lagrangian then reads
\begin{eqnarray}
  \mathcal L&=&-
  \{\4 F_{\mu\nu}^2+\2(D_\mu S)^2+\2(D_\mu P)^2+\2
  g^2(S\times P)^2\}\nonumber\\
  \label{eq:L4d} 
  &&-\{
  \bar\psi\gamma^\mu D_\mu\psi +ig\bar\psi (S\times \psi)
+g\bar\psi\gamma_5(P\times\psi)\}.
\end{eqnarray}
In the trivial sector we choose
the symmetry-breaking Higgs field to be $S^3$ with background
$S^a\equiv A_6^a=v \delta^a_3$;
in the nontrivial sector we consider a static monopole background with only
$A_j^a$ ($j=1,2,3$) and $S^a$ nonvanishing.
The classical Hamiltonian density can be written in BPS form
\be\label{Ham}
\mathcal H=\4(F_{ij}^a-\epsilon_{ijk}D_k S^a)^2+\2\6_k(\epsilon_{ijk}
F_{ij}^a S^a).
\ee
(One can also write $\mathcal H$ in similar form if $A_0$ is nonvanishing
(dyons) \cite{Bogomolny:1976de,Coleman:1976uk,Weinberg:1979ma,Chalmers:1996ya}, but we shall not need this extension.)
Thus the classical BPS equations for the monopole read 
\be\label{BPSeq}
F_{ij}^a-\epsilon_{ijk}D_k S^a=0.
\ee
One can interpret this equation as a selfduality
condition for $F_{MN}$
with $M,N=1,2,3,6$
and $\epsilon_{ijk6}=\epsilon_{ijk}$. 
The anti-monopole is obtained by reversing the
sign of $S^a$ in eqs.~(\ref{BPSeq}) and (\ref{Ham}).

To solve the BPS equation one may set
$A_i^a=\epsilon_{aij}{x^j\0gr^2}f(r)$ and
$S^a=-m{x^a\0gr}h(r)$ with $m=gv$ and
impose the boundary conditions $f(r)\to1$ and $h(r)\to1$ for $r\to\infty$.
This leads to two coupled
first-order ordinary differential equations for $f(r)$ and $h(r)$,
\be
-2f+f^2+r^2mh'=0,\qquad f'=mh(1-f),
\ee
whose solution was obtained by Prasad and Sommerfield \cite{Prasad:1975kr}
\be
f=1-mr/\sinh(mr),\qquad h=\coth(mr)-(mr)^{-1}.
\ee
For us only the asymptotic values of the background fields
will be needed. These are given by
\bea\label{asympt}
A_i^a \to \epsilon_{aij}{1\0g}{\hat x^j\0r},\qquad
F_{ij}^a\to -\epsilon_{ijk}{1\0g}{\hat x^a \hat x^k\0r^2},\nn
S^a\to -\delta_i^a \hat x^i v(1-{1\0mr}),\qquad 
D_iS^a \to -{1\0g}{\hat x^i \hat x^a \0 r^2},
\eea
where $\hat x^i \equiv x^i/r$.
Substituting these expressions into the surface term in
the Hamiltonian (\ref{Ham})
yields the classical mass 
\be
M_{\rm cl.}=4\pi m/g^2.
\ee
Note that the Hamiltonian (\ref{Ham})
corresponds to the standard expression for
the gravitational stress tensor 
associated with the Lagrangian density
in (\ref{eq:L6d}).
As we shall discuss further below,
one can also define an improved stress tensor, 
and then one obtains in fact a different value
for the classical mass.

The action is invariant under susy transformations with a complex
chiral spinor $\eta=\left(\epsilon\atop 0\right)$
\be
\delta A_M^a=\bar\lambda^a \Gamma_M \eta-\bar\eta \Gamma_M \lambda^a ,\quad
\delta \lambda^a=\2 F^a_{MN}\Gamma^M \Gamma^N \eta,
\quad
\delta \5\lambda^a=-\2 \5\eta F^a_{MN}\Gamma^M \Gamma^N .
\ee
The susy current as obtained from the Noether method applied
to transformations with $\bar\eta$ reads
\be\label{jM}
j^M=\2\Gamma^P\Gamma^Q F_{PQ}^a \Gamma^M \lambda^a
\equiv \2 \Gamma^{PQ}F_{PQ}^a \Gamma^M \lambda^a.
\ee
It is gauge invariant and conserved on-shell.
Its susy variation yields\footnote{We have displayed the result for
$\2 \delta j^M$ instead of $\delta j^M$ in order to obtain the
stress tensor with unit normalization on the right-hand side.}
\bea\label{djN2}
\2 \delta j^M&=&\8{1\08} (\Gamma^{PQMRS} \eta) F_{PQ} F_{RS}
+ \Gamma^N\eta(F^{MP} F_{NP} - \4 \delta^M_N F^{RS} F_{RS})\nn
&&-((D^N\5\lambda)\Gamma^M\lambda)\Gamma_N\eta
+(\lambda^T C_{6}\Gamma^M D_N\lambda)\Gamma^N \tilde\eta
\eea
where $\tilde\eta=C_{6}^{-1}\bar\eta^T$ and
we have dropped terms proportional to the equation of motion
$\Gamma^M D_M \lambda=0$.
The terms with $\Gamma^N\eta$ should give the stress tensor $T_{MN}$.
For the fermions one finds $-(D_N\5\lambda)\Gamma_M\lambda$,
which seems to differ from the gravitational stress tensor\footnote{To
obtain the gravitational stress tensor, one may replace $\Gamma^A \Gamma^{BC}$
(with $A,B,C$ flat vector indices) in the term
$-e(\5\lambda\Gamma^A \Gamma^{BC}\lambda)e_A^M\4\omega_{MBC}$
in the action by one-half the commutator because the
anticommutator only yields terms with two or more
vielbein fluctuation fields $e_A^M-\delta_A^M$.
One must then evaluate the variation of $ee_A^M\omega_{MB}{}^A$ which is
easiest obtained by using the vielbein postulate
$\6_M(ee_B^M)+ee_A^M\omega_{MB}{}^A=0$. This
yields $\delta S=\int[-\5\lambda \Gamma^A D_M^{\rm (YM)}\lambda
+\2\6_M(\5\lambda \Gamma^A \lambda)]\delta(ee_A^M)d^4x$
up to terms quadratic in $e_A^M-\delta_A^M$,
where $D_M^{\rm (YM)}$ is the flat-space Yang-Mills covariant
derivative.\label{fn6}}
$-(\delta S/\delta e_A{}^M)e_{AN}=\2\5\lambda \Gamma_N \olrD_M \lambda$.
However, on-shell the gravitational stress tensor is symmetric
as a consequence of the local Lorentz invariance of the action
in curved space, and $-(D_N\5\lambda)\Gamma_M\lambda$ differs
from $\2\5\lambda \Gamma_N \olrD_M \lambda$ by the total
derivative $-\2\6_N(\5\lambda \Gamma_M\lambda)$. This total
derivative is separately conserved with respect to the index $M$,
and does not contribute to the translation generators,
$\2\int\6_N(\5\lambda\Gamma^0\lambda)d\vec x=0$.\footnote{For
$N=j$ this is clear, but it also holds for $N=0$ if one uses
the Dirac equation.}
Thus the susy algebra does not depend on which stress tensor one uses,
and for the energy it does not matter whether one uses
$\int T_{00}=-\int (D_0\5\lambda)\Gamma_0 \lambda$ or
$\int T_{00}=\2\int(\5\lambda\Gamma_0\olrD_0 \lambda)$.

Integrating over 3-space, the term with $\tilde\eta$ vanishes
since $\lambda^T C_{6}\Gamma^M D_N\lambda=\2
\6_N(\lambda^T C_{6}\Gamma^M \lambda)$,
while the terms with $\eta$ yield the $N=2$ susy algebra
\be
\label{n2sag}
\8{i\02}\{ Q^\alpha,\5 Q_\beta\}
=(\gamma^\mu)^\alpha{}_\beta P_\mu -(\gamma_5)^\alpha{}_\beta \hat U
+i\delta^\alpha_\beta \hat V.
\ee
Here $P_\mu=\int T^0{}_\mu d^3x$ and
$\hat U$ and $\hat V$ are the two real (or one complex) central charges
of the $N=2$ algebra, given explicitly by\footnote{We have
added hats to $\hat U$ and $\hat V$ to distinguish them from
the $N=4$ case, where we shall introduce the
four central charges $U,\tilde U,V,\tilde V$.}
\bea
\label{UV}
\hat U&=&\int [\2\epsilon^{ijk}\6_k (F_{ij}^a S^a) 
+\6_i(P^a F^a_{i0}) ] d^3x ,\\
\label{UVb}
\hat V&=&\int [\2\epsilon^{ijk}\6_k (F_{ij}^a P^a)
- \6_i(S^a F^a_{i0}) ]
d^3x.
\eea

For a monopole background $\hat U=\int \2\epsilon^{ijk}\6_k (F_{ij}^a S^a)
d^3x$ which saturates the BPS bound classically, see (\ref{Ham}). 
Performing the integration over 3-space with a monopole
background yields $\hat U_{\rm cl.}={4\pi m/g^2}$.

It is well-known that one can add improvement terms 
$\Delta T_{\mu\nu}^{\rm impr}$ to the
stress tensor for 
scalar fields so that the improved stress tensor
becomes traceless on-shell
\be\label{TmnimprS}
T_{\mu\nu}(S)+
\Delta T_{\mu\nu}^{\rm impr}(S)=
D_\mu S D_\nu S - \2\eta_{\mu\nu} (D_\rho S)^2
-\8{1\06} (\6_\mu \6_\nu - \eta_{\mu\nu}\6^2) S^2
\ee
and similarly for $P$. (One may verify that the complete classical
stress tensor for $A_\mu, P, S, \psi$ is indeed traceless when
the full nonlinear field equations for $P$ and $S$ are satisfied, 
even though some of the fields are massive due to the Higgs effect.)

This corresponds to the possibility of introducing
improvement terms to the susy current so that the improved
susy current satisfies $\gamma^\mu j_\mu=0$ (``gamma-tracelessness'')
in 3+1 dimensions. In 5+1-dimensional notation
\mpar{Is there an anomaly in the impr. term?}
\be
j^\mu_{\rm impr}\equiv j^\mu+\Delta j^\mu_{\rm impr}=
\2 \Gamma^{PQ} F_{PQ} \Gamma^\mu \lambda
-\8{2\03} \Gamma^{\mu\nu} \6_\nu (A_{\mathcal J} \Gamma^{\mathcal J} \lambda)
\ee
where the index $\mathcal J$ runs over $\mathcal J=5,6$ in
the $N=2$ model (in the $N=4$ model it will run from 5 to 10).
In fact, the improved currents on the one hand, and the
improvement terms on the other hand, form separate $N=2$ multiplets.
Notice that the improvement terms are perfectly gauge invariant
quantities in 3+1 dimensions, but not in the higher-dimensional
ancestor model because there $A_{\mathcal{J}}$ are components of a gauge field. 

A susy variation of the improvement term by itself yields the following
result
\bea\label{dDjimpr2}
\2\delta(\Delta j^\mu_{\rm impr})&=&-\8{1\06}\gamma^{\mu\nu}\6_\nu
(A_\J \Gamma^\J F_{MN} \Gamma^{MN}\eta) \nn
&&-\8{1\03}\gamma^{\mu\nu}\6_\nu(\Gamma^\J \lambda \{
\5\lambda \Gamma_\J \eta - \5\eta \Gamma_\J \lambda \} ).
\eea
The bosonic terms in the $\mu=0$ component of this expression yield
\bea
&&-{1\03}\gamma^0\gamma^j\6_j(P\gamma_5+iS)
(\2F_{\rho\sigma}\gamma^{\rho\sigma}+D_\rho P\gamma^\rho \gamma_5
+D_\rho S i \gamma^\rho)\eta\nn
&=&-{1\03}\gamma^0\gamma^j\6_j
[(P\gamma_5+iS)\2F_{\rho\sigma}\gamma^{\rho\sigma}-\2\7D(S^2+P^2)
-i\gamma^\rho \gamma_5 P \olrD_\rho S]\eta\qquad.
\eea
The terms with $S^2+P^2$ yield the improvement terms of the stress
tensor, and the terms with $(\gamma_5)^\alpha{}_\beta$ and
$i\delta^\alpha{}_\beta$ are equal to $-{1\03}$ times the results
in (\ref{n2sag}), (\ref{UV}), and (\ref{UVb}).
There are, however, further terms with a different spinor structure
and while the $N=4$ case has been worked
out at the linearized level
in Ref.~\cite{Bergshoeff:1980is,Bergshoeff:1982av}, 
the complete multiplet of $N=2$ improvement terms seems
not to be known. There are also fermionic terms, which read
after a Fierz rearrangement\be
{1\012}\gamma^{\mu\nu}\6_\nu[(\5\lambda O^I \lambda)\Gamma_\J O_I
\Gamma^\J \eta + (\lambda^T C_{6} O^I \lambda)\Gamma_\J O_I
\Gamma^\J \tilde\eta]
\ee
where we used $\5\eta\Gamma_\J\lambda=-\lambda^T\Gamma_\J^T\5\eta^T
=(\lambda^T C_{6})\Gamma_\J (C_{6}^{-1} \5\eta^T)
=\lambda^T C_{6} \Gamma_\J \tilde \eta $.
Since $\lambda^T C_{6} O^I \lambda$ is only nonvanishing for
$O^I=\Gamma^M,\Gamma^{M_1\cdots M_5}$, we find no fermionic 
improvement terms for the central charges $\hat U$ and $\hat V$.

As we have seen,
the improvement terms for the central charges $\hat U$ and $\hat V$
are proportional to the unimproved ones,
\be
\Delta \hat U^{\rm impr}=-{1\03} \hat U.
\ee
This shows that using an improved supercurrent multiplet
reduces the classical value for the monopole mass
and central charge to $2/3$ of its conventional (unimproved)
value. (For the mass this can be verified explicitly by evaluating the
improvement term (\ref{TmnimprS}) using the asymptotic
results for the scalar field of eq.\ (\ref{asympt}).)

\subsection{N=4}

We now turn to the $N=4$ super Yang-Mills theory with monopoles.
Most aspects are the same as in the $N=2$ case, but for the
improvement terms there are important differences.

The $N=4$ model is most easily obtained by dimensional reduction
from 9+1 dimensions
\cite{Brink:1977bc,Gliozzi:1976qd}. The Lagrangian is given by 
\bea
  \label{eq:L10d}
    \mathcal L
&=&- \4 F_{MN}^2 - \2 \bar{\lambda}\Gamma^M D_M\lambda\nn
&=&-\4 F_{\mu\nu}^2-\2(D_\mu S_j)^2-\2(D_\mu P_j)^2
-\2 \5\lambda^I \7D \lambda^I + \mbox{interactions}
\eea
where the factor $\2$ is due to the fact that 
the fermionic field $\lambda$ is now a Majorana-Weyl spinor, and
the indices $M,N$ now run over $0,\ldots3,5,\ldots 10$.
After dimensional reduction there are three adjoint scalar and pseudoscalar
fields, indexed by $j=1,2,3$, and four adjoint Majorana fields indexed
by $I=1,\ldots,4$, but it will be often convenient to keep
the 10-dimensional notation.
The action is invariant under $\delta A_M=-\bar\epsilon \Gamma_M \lambda
=\5\lambda \Gamma_M \epsilon$
and $\delta \lambda = \2 F_{MN} \Gamma^{MN} \epsilon$.

The susy current is 
again formally given by (\ref{jM})
but now varies under susy into
\bea\label{dj}
\2 \delta j^M(x) &=&
(F^{MP} F_{NP} - \4 \delta^M_N F^{RS} F_{RS} + \2 \bar\lambda
\Gamma^M D_N \lambda) \Gamma^N \epsilon \nn
&&+{1\016}(\bar\lambda \Gamma^M \Gamma^{PQ} D^R \lambda) \Gamma_{PQR}\epsilon\nn
&&+{1\08}F_{NP}F_{QR} \Gamma^{NPMQR} \epsilon
\eea
The term with $\Gamma_{PQR}$ vanishes after integration over $x$.\footnote{%
One can rewrite $(\bar\lambda\Gamma^M\Gamma^{PQ}D^R\lambda)\Gamma_{PQR}\epsilon$
as $\5\lambda\{ {1\06}[\Gamma^{MPQR},\Gamma^N]
D_N\lambda+{1\03}\5\lambda \Gamma^{PQR} D^M \lambda
+\2\5\lambda [\Gamma^{QR},\Gamma^N]D_N \lambda \} \Gamma^M{}_{QR}\epsilon$
and $\int \5\lambda \Gamma^{PQR}D^M\lambda$ vanishes
since $\5\lambda \Gamma^{PQR}D^M\lambda=(D^M\5\lambda) \Gamma^{PQR}\lambda$
is a total derivative of fermionic terms.}

For the purpose of dimensional reduction, the 16-component Majorana-Weyl
spinor $\lambda^a$ is written as $(\lambda^{\alpha Ia},0)$, where
$\alpha=1,\ldots,4$ is
the 4-dimensional spinor index, 
$I=1,\ldots,4$ is the rigid SU(4) index, and $a$ is the adjoint
SU(2) colour index.
For the gamma matrices we use the representation\footnote{
Majorana spinors satisfy $\lambda^TC_{10}=\lambda^\dagger
i\Gamma^0$  in 9+1 dimensions
and $(\lambda^I)^TC_4=(\lambda^I)^\dagger i \gamma^0$
in 3+1 dimensions.}
\bea\label{Gammas}
\Gamma_\mu &=& \gamma_\mu \otimes {\bf 1} \otimes \sigma_2, 
\quad \mu=0,1,2,3,\nn
\Gamma_{4+j}&=& {\bf 1} \otimes \alpha_j \otimes \sigma_1, \nn
\Gamma_{7+j}&=& \gamma_5 \otimes \beta_j \otimes \sigma_2,
\quad j=1,2,3, \nn
\Gamma_{11}&=&{\bf 1} \otimes {\bf 1} \otimes \sigma_3,
\qquad C_{10}=-iC_4\otimes {\bf 1} \otimes \sigma_1.
\eea
The $\alpha_j$ and $\beta_j$ are the six generators of SO(4) = SO(3)$\times$SO(3)
in the representation of purely imaginary antisymmetric $4\times4$ 
matrices \cite{'tHooft:1976fv,Figueroa:EDC},
self-dual and anti-self-dual, respectively, and
satisfying $\{\alpha_i,\alpha_j\}=\{\beta_i,\beta_j\}=2\delta_{ij}$,
$[\alpha_i,\alpha_j]=2i\epsilon_{ijk}\alpha_k$,
$[\beta_i,\beta_j]=2i\epsilon_{ijk}\beta_k$,
and $[\alpha_j,\beta_k]=0$. 
An explicit representation is given by
\begin{eqnarray}
  \label{eq:abrep}
  &&\alpha_1=\sigma_1\otimes\sigma_2\ ,\ \alpha_2=-\sigma_3\otimes\sigma_2\ ,\
  \alpha_3=\sigma_2\otimes\unit \\\nonumber
  &&\beta_1=\sigma_2\otimes\sigma_1\ ,\ \beta_2=\unit\otimes\sigma_2\ ,\ 
  \beta_3=\sigma_2\otimes\sigma_3.
\end{eqnarray}

Upon
reduction to 3+1 dimensions, 
the spin zero fields $A_\J$ with $\J=5,\ldots,10$
split into three scalars $S_j=A_{4+j}$ and three pseudoscalars
$P_j=A_{7+j}$. 

The 12 real central charges appear as\footnote{The 
complex matrix $Z^{IJ}=-Z^{JI}$ of central charges 
contains 6 complex (12 real) elements, as shown in (\ref{QQ}).
The magnetic charge $U_1$ and the electric charge $\tilde V_1$
only appear in the combination $Z^{IJ}=(iU_1+\tilde V_1)(\alpha^1)^{IJ}
+\ldots$ in the left-handed sector.
By a unitary
transformation one can block-diagonalize $Z^{IJ}$, with two real antisymmetric
2$\times$2 matrices along the diagonal.
The $N=4$ action has a
rigid $R$ symmetry group SU(4) (not U(4) \cite{Sohnius:1985qm}).
To exhibit this SU(4), the spin zero fields
are combined into $M^{IJ}=(\alpha^j)^{IJ}S_j +i (\beta^j)^{IJ}P_j$, 
but if $S_1^a=v\delta^a_3$, there is a
central charge acting on excitations
and $R$ is broken down to the manifest stability subgroup of
$\<M^{23}\>=\<M^{14}\>=S_1$,
which is USp(4). (Note that the central charge and all susy
generators vanish on the trivial spontaneously broken vacuum).}
\bea\label{QQ}
\2\{ Q^{\alpha I},Q^{\beta J} \} &=&
\delta^{IJ} (\gamma^\mu C^{-1})^{\alpha\beta} P_\mu \nn
&+&\!\!\!i(\gamma_5 C^{-1})^{\alpha\beta} (\alpha^j)^{IJ} \intx U_j
-(C^{-1})^{\alpha\beta} (\beta^j)^{IJ} \intx V_j \nn
&+&\!\!\!(C^{-1})^{\alpha\beta} (\alpha^j)^{IJ} \intx \tilde V_j
+i(\gamma_5 C^{-1})^{\alpha\beta} (\beta^j)^{IJ} \intx \tilde U_j\,,\quad
\eea
where $U_j$ and $V_j$ are due to the five-gamma term in (\ref{dj}),
and $\tilde U_j$ and $\tilde V_j$ due to the one-gamma terms.
The indices $I$ and $J$ are lowered and raised by the charge conjugation
matrix in this space, which is $\delta^{IJ}$, see (\ref{Gammas}).

The $N=2$ monopole (and dyon) can be embedded 
into the $N=4$ model by selecting e.g.\ $j=1$
and one then has ($S=S_1$, $P=P_1$, $U=U_1$ etc.)
\footnote{To obtain the total derivatives in (\ref{UUVV}),
one needs to use Bianchi identities in the case of $U$ and $V$, and
equations of motion in the case of $\tilde U$ and $\tilde V$.}
\bea\label{UUVV}
&&U=\6_i(S^a \2 \epsilon^{ijk} F_{jk}^a),
\qquad \tilde U=\6_i(P^a F_{0i}^a) \nn
&&V=\6_i(P^a \2 \epsilon^{ijk} F_{jk}^a),
\qquad \tilde V=\6_i(S^a F_{0i}^a).
\eea
The monopole background fields are again given by (\ref{asympt}).

Let us now discuss the improvement terms. For the stress tensor
they are given by
\be\label{Tmunuimpr}
\Delta T_{\mu\nu}^{\rm impr}=-{1\06}(\6_\mu \6_\nu - \eta_{\mu\nu}\6^2)
(A_\J A^\J)
\ee
where $\J=5,\ldots,10$. Also there is an improvement term in the supercurrent
\be
\2\Delta j^\mu_{\rm impr}=
-{1\03} \Gamma^{\mu\nu} \6_\nu (A_{\mathcal J} \Gamma^{\mathcal J} \lambda)
\ee
The susy variation of $\Delta j^\mu_{\rm impr}$ is simpler to
evaluate than in the $N=2$ case because both $\lambda$ and $\epsilon$
are Majorana spinors. One finds
\be
\2\label{dDjimpr4}
\delta(\Delta j^\mu_{\rm impr})=-{1\03}\Gamma^{\mu\nu}\6_\nu
\left[ A_\J \Gamma^\J\2F_{MN}\Gamma^{MN}\epsilon
-(\bar\epsilon \Gamma_\J\lambda)\Gamma^\J \lambda\right]
\ee

The bosonic terms read more explicitly
\begin{equation}
  \label{eq:imZbos}
  -\frac{1}{6}\gamma^{\mu\nu}\partial_\nu(\alpha^j\otimes\sigma_1 S_j + 
    \gamma_5\otimes\beta^j\otimes\sigma_2 P_j) 
    (\gamma^{\rho\sigma}F_{\rho\sigma}
-2i\gamma^\rho\alpha^l\otimes\sigma_3 D_\rho S_l
+2\gamma^\rho\gamma_5\beta^l D_\rho P_l)\epsilon.
\end{equation}
They yield again $-{1\03}$ times the unimproved central charges.
From the fermionic terms one obtains for $\mu=0$ after a Fierz
rearrangement
\be
{1\03}\Gamma^{0j}\6_j [ {1\024}(\5\lambda \Gamma^{PQR}\lambda) ]
\Gamma_{PQR}\epsilon=
-{1\03}\epsilon^{jkl}\6_j [ {i\08}\5\lambda\gamma_{kl}\alpha^1\lambda ]
(i\gamma_5\alpha^1\epsilon)
+\ldots
\ee
Thus one obtains the following
improvement terms for the central charges
\be\label{UimprN4}
\Delta U^{\rm impr}_1=-{1\03}\left[U_1+\int {i\08}\6_i (\epsilon^{ijk}
\bar \lambda \alpha^1 \gamma_{jk} \lambda) d^3x \right]
\ee
etc.
Classically, there is no contribution from the fermionic terms, so
like in the $N=2$ the use of improved susy currents and stress tensors
leads to a mass and central charge of a monopole equal to 2/3 of
their conventional (unimproved) value.

\section{Fluctuations about the monopole background and
  renormalization}

To perform quantum calculations one needs to add a gauge fixing term and ghosts to
the classical Lagrangian. Usually fermions do not enter the gauge fixing condition.
Therefore, as for the bosonic Lagrangian, we formally don't have to differentiate 
between the  $N=2$ and the $N=4$ model in this respect.
The background-covariant
Feynman-$R_\xi$ gauge at $\xi=1$, implemented in the higher dimensional
space,  has turned out to be the most convenient gauge in the quantum theory of
solitons {\cite{Rebhan:2003bu,Rebhan:2004vn}}. 
For general $\xi$ the BRS-exact gauge fixing 
Lagrangian including ghosts is given by\footnote{Note that for $\xi\neq 1$ the 
kinetic terms of the quantum fluctuations are not diagonal; for that one would
need the non-background covariant gauge of the form 
$-\frac{1}{2\xi}[\partial_\mu a^\mu+\xi(\ldots)]^2$.}
\begin{equation}
  \label{Lgf}
  {\cal{L}}_{\mathrm{gf+gh}}= -\frac{1}{2\xi}[D_M(A)a^M]^2 
  - [D^M(A) b]\ D_M(A+a) c\ ,
\end{equation}
where $a_M$ are the quantum fields and $A_M$ the background fields.

\subsection{Fluctuation equations and propagators}
\label{sec:fluct}

For the calculation of one-loop corrections we need the
terms in the action that are
quadratic in quantum fluctuations about a (monopole) background.
Propagators and the fluctuation equations are then obtained from this 
quadratic 
Lagrangian. Expanding the bosonic Lagrangian in (\ref{eq:L4d}) and the gauge 
fixing Lagrangian (\ref{Lgf}) around the solution of (\ref{BPSeq}),
$  A_\mu\rightarrow \bg{A}_\mu+a_\mu,\ S\rightarrow \bg{S}+s,\ P\rightarrow p\ ,$
the quadratic Lagrangian reads for $\xi=1$ and $N=2$
\begin{eqnarray}
  \label{eq:lbgh2}
  \CL^{(2)}_{\mathrm{bos+gh}}=&-&\hal\ a_{\9 m}
    \[(\6_0^2-\6_5^2-D_{\9\ell}^2)\delta_{\9 m\9 n}
       -2 g \bg{F}_{\9 m \9 n} \times \]a_{\9 n} \\ \nonumber
    &-& \hal\ p(\6_0^2-\6_5^2-D_{\9 m}^2)p \\ \nonumber
    &+& \hal\ a_0(\6_0^2-\6_5^2-D_{\9 m}^2)a_0 
       - b(\6_0^2-\6_5^2-D_{\9 m}^2)c \ ,
\end{eqnarray}
where we kept the derivative  w.r.t. one of the extra dimensions, $\6_5$,  
for the purpose of 
regularization. The quantum fields $a_i$ and  $s$ are
combined in a quartet 
$a_{\9 m}=(a_i,s)$. The background field strength $\bg{F}_{\9 m \9 n}$
and the background covariant derivative $D_{\9 m}$ are defined 
correspondingly.\footnote{The various boundary terms 
that we omitted in (\ref{eq:lbgh2}) may give contributions  at the boundary 
of space-time but 
they are not included in the definition of the propagators.} 
The $N=4$ model has four
additional (pseudo)scalars $q_I$ with the same quadratic Lagrangian as for 
the pseudoscalar $p$.

The linearized field equations, i.e the fluctuation equations, 
are given by
\begin{eqnarray}
\label{eqn:bosfluct}
 && \[(\6_0^2-\6_5^2-D_{\9\ell}^2)\delta_{\9 m\9 n}
  -2 g \bg{F}_{\9 m \9 n} \times \]a_{\9 n}=0\ ,\nn
 && (\6_0^2-\6_5^2-D_{\9 m}^2)(p,a_0,b,c)=0\ .
\end{eqnarray}
The propagators are given correspondingly by
\begin{eqnarray}
  \label{eq:props}
  \<a^b_{\9m}(x)a^c_{\9n}(y)\>
  &=&i\<x|[((\6_5^2-\6_0^2)\delta^{bc}+(D_{\9k}^2)^{bc})\delta_{\9m\9n}
  +2g\epsilon^{bac}\bg{F}^a_{\9m\9n}]^{-1}|y\> \ ,\\ 
  \<p^b(x)p^c(y)\>&=& i\<x|[(\6_5^2-\6_0^2)\delta^{bc}+(D_{\9k}^2)^{bc}]^{-1}|y\> \ .
\end{eqnarray}
Note that the propagators for the ghosts and the component $a_0$ have an opposite 
sign, i.e.
\begin{equation}
  \label{eq:aghprop}
  \<p^b(x)p^c(y)\>= -\<a_0^b(x)a_0^c(y)\>= -\<b^b(x)c^c(y)\>. 
\end{equation}

The fermionic fluctuations are independent of the gauge condition
chosen for the bosonic fields. 
For the $N=2$ model they are obtained from the dimensional reduction 
of $\Gamma^M D_M\lambda=0$, which using (\ref{eq:6dgam}) gives
\begin{equation}
  \label{eq:n2psi}
  (\gamma^\mu D_\mu+\gamma_5\6_5)\psi+ig\bg{S}\times\psi=0\ .
\end{equation}
In order to obtain a
diagonal action for the two complex 2-component spinors into
which $\psi$ of the $N=2$ model decomposes,\footnote{The factor $i$ in front 
of $\psi_-$ is inserted in order that
the field operators for $\psi_+$ and $\psi_-$ have the same form 
as for the kink and the vortex that we shall discuss in section \ref{sec:indexth}.}
$\psi=\left( \psi_+ \atop
i \psi_-\right)$, we use an unconventional representation for
the 4-dimensional Dirac matrices,
\be
\label{4dgamma}
\gamma^k=\pmatrix{ \sigma^k & 0 \cr 0 & -\sigma^k },\quad
\gamma^0=\pmatrix{ 0 & i \cr i & 0 },\quad
\gamma_5=\pmatrix{ 0 & -i \cr i & 0 }.
\ee

One finds then the following two field equations for $\psi_+$ and $\psi_-$
in a monopole background
\begin{equation}
  \label{eq:f1}
  {\7{\sf D}} \psi_+ = (\6_0-\6_5)\psi_-\quad ,\quad 
  \bar{\7{\sf D}} \psi_- = (\6_0+\6_5)\psi_+\ ,
\end{equation}
where the operators in (\ref{eq:f1}) are defined by
\begin{equation}
  \label{eq:dslash}
  {\7{\sf D}}:=\sigma^{\9 m}D_{\9 m}=\sigma^k D_k+ig\bg{S}\times\quad ,\quad 
  \bar{\7{\sf D}}:=\bar\sigma^{\9 m}D_{\9 m}=\sigma^k D_k-ig\bg{S}\times\ .
\end{equation}
Here we have introduced the Euclidean sigma matrices 
$\sigma^{\9 m}=(\sigma^k,i\unit)$ and 
$\bar\sigma^{\9 m}=(\sigma^k,-i\unit)$.
Iteration yields
\be\label{fermeoms}
\bar{\7{\sf D}}{\7{\sf D}}\ \psi_+=(\6_0^2-\6_5^2)\psi_+\quad ,\quad  
{\7{\sf D}}\bar{\7{\sf D}}\ \psi_-=(\6_0^2-\6_5^2)\psi_-\ .
\ee
In terms of the self-dual background $\bg{F}_{\9 m\9 n}$  one obtains
\begin{equation}
  \label{eq:d2}
  \bar{\7{\sf D}}{\7{\sf D}}=
    D^2_{\9 m}+\hal\ \bar\sigma^{\9 m \9 n}g\bg{F}_{\9 m\9 n}\quad ,\quad 
  {\7{\sf D}}\bar{\7{\sf D}}=D^2_{\9 m}\ , 
\end{equation}
where we have used the self-dual 
Lorentz generators
$\bar\sigma^{{\underline m}{\underline n}}=\2(\bar\sigma^{\underline m}
\sigma^{\underline n}-\bar\sigma^{\underline n}\sigma^{\underline m})$. 

Introducing spinor notation for  the quartet $a_{\9 m}=(a_i,s)$, i.e. 
$\slasha{a}=\sigma^{\9 m}a_{\9 m}$ and 
$\bar\slasha{a}=\bar\sigma^{\9 m}a_{\9 m}$,   
one can rewrite the bosonic part of the Lagrangian
quadratic in quantum fields in spinor notation as follows
\begin{eqnarray}
  \label{eq:L2bos}
  \CL^{(2)}_{\rm bos.}&=&
\4\tr \biggl[\7a(\7{\5{\sf D}}{\7{\sf D}}+\6_5^2-\6_0^2)\7{\5a}-
a_0({\7{\sf D}}\7{\5{\sf D}} 
+\6_5^2-\6_0^2)a_0\nn&&+p({\7{\sf D}}\7{\5{\sf D}} +\6_5^2-\6_0^2)p
+2b({\7{\sf D}}\7{\5{\sf D}} +\6_5^2-\6_0^2)c \biggr]\ .
\end{eqnarray}
Thus the four bosonic fields $\7{a\,}$
satisfy the same fluctuation equations as $\psi_+$, 
whereas the quartet\footnote{This quartet is different
from the usual Kugo-Ojima quartet \cite{Kugo:1978zq}. The pseudo Goldstone
fields are given by $s_1^{1,2}$. Note also that these are massive, whereas
the Higgs field $s_1^3$ is massless.}
$(a_0,p,b,c)$   satisfies the same
equations as $\psi_-$.
In the $N=4$ case the additional scalars $q_I$ of that theory
have the same fluctuation equations as this quartet.

Although the gauge fixing term in (\ref{Lgf}) is not
susy invariant, for $\xi=1$ it gives rise to a 
(quantum mechanically) supersymmetric
set of fluctuation equations. The factorization property (\ref{eq:d2})  
implies that the
two operators ${\7{\sf D}}\bar{\7{\sf D}}$ and $\bar{\7{\sf D}}{\7{\sf D}}$ are 
isospectral, except for zero modes. The respective normalized eigen-modes of
$\psi_+$ and $\psi_-$ with mode 
energy $\omega^2_k=k^2+m^2$,
\begin{equation}
  \label{eq:modes}
  -\bar{\7{\sf D}}{\7{\sf D}}\ \chi^+_k = \omega^2_k\ \chi^+_k \quad ,\quad
  -{\7{\sf D}}\bar{\7{\sf D}}\ \chi^-_k = \omega^2_k\ \chi^-_k\ ,
\end{equation}
are related to each other as follows:
\begin{equation}
  \label{eq:susqm1}
  \chi^-_k=\frac{1}{\omega_k}\ {\7{\sf D}}\ \chi^+_k \ .
\end{equation}
This susy quantum mechanical structure will allow us to use
index theorems for the calculation of the difference
of the spectral densities associated with the
operators ${\7{\sf D}}\7{\5{\sf D}}$ and $\7{\5{\sf D}} \7{\sf D}$.

In $4+\epsilon$ dimensions
the fermionic quantum field has the following mode expansion
\bea
\label{eq:fqf}
\psi(x)\!&=&\!
{ \psi_+ \choose i\psi_- } =
\int\frac{d^{\epsilon}\ell}{(2\pi)^{\epsilon/2}}
\intsum {d^3k\0(2\pi)^{3/2}}{1\0\sqrt{2\omega}}
\biggl\{ b_{kl} e^{-i(\omega t - \ell x^5)}
{\sqrt{\omega+\ell}\; \chi_k^+ \choose - \sqrt{\omega-\ell} \;\chi_k^- }\nonumber\\
&&\qquad\qquad +d_{kl}^\dagger e^{i (\omega t - \ell x^5)}
{ \sqrt{\omega+\ell} \;\chi_k^+ \choose  \sqrt{\omega-\ell} \;\chi_k^- }
\biggr\}\ +\mathrm{zero\  modes}\ ,
\eea
where $\chi_k^{\pm}$ depend on $x^1$, $x^2$, $x^3$ and the 
regularized mode energies are given by
\begin{equation}
  \label{eq:omreg}
  \omega^2=\omega_k^2+\ell^2\ .
\end{equation}
The bosonic fields $a_{\9 m}$ in spinor notation
and the quartet $a_0,p,b,c$ have
an analogous mode expansions in terms of  
$\chi_k^-$ and $\chi_k^+$, respectively.  For example
\begin{equation}
  \label{eq:a0}
  a_0(x)= \int\frac{d^{\epsilon}\ell}{(2\pi)^{\epsilon/2}}
\intsum {d^3k\0(2\pi)^{3/2}}{1\0\sqrt{2\omega}}
\left(a_{k\ell} e^{-i(\omega t - \ell x^5)} \chi_k^-+\textrm{h. c.}\right ).
\end{equation}

For the $N=4$ model the dimensional reduction using (\ref{Gammas})
gives the following fluctuation equations for the fermions in the monopole 
background
\begin{equation}
  \label{eq:n4psi}
  \gamma^\mu D_\mu\psi_I-ig\bg{S}\times\alpha^1_{IJ}\psi_J=0\ ,
\end{equation}
where $\alpha^1_{IJ}$ was given in (\ref{eq:abrep}) and we have omitted
the regulating extra dimension since the $N=4$ model is non-anomalous. 
The $\psi_I$ are four 4-component $D=3+1$ Majorana spinors.
The mode expansions of these Majorana spinors contains an extra
factor $1/\sqrt2$ compared to (\ref{eq:fqf}).
One can decouple the equation 
in the $SU(4)$-space by introducing complex linear combinations
$\psi_{(I,I\!I)}=
{1\0\sqrt2}(\psi_1\pm i\psi_4)$ and
$\psi_{(I\!I\!I,I\!V)}={1\0\sqrt2}(\psi_2\pm i\psi_3)$.  
The resulting equations are
\bea
  \label{eq:n4psi2}
&&\gamma^\mu D_\mu\psi_{(I,I\!I\!I)}+ig\bg{S}\times\psi_{(I,I\!I\!I)}=0\ ,\nn
&&\gamma^\mu D_\mu\psi_{(I\!I,I\!V)}-ig\bg{S}\times\psi_{(I\!I,I\!V)}=0.
\eea
The first two equations
coincide with the $N=2$ equation (\ref{eq:n2psi}).
(Setting $\psi_1=i\psi_4$ and $\psi_2=i\psi_3$, together
with $\epsilon_1=i\epsilon_4$ and $\epsilon_2=i\epsilon_3$
as well as $P_2=P_3=S_2=S_3=0$, is a consistent truncation from
$N=4$ to $N=2$.)

\subsection{One-loop counterterms}
\label{sec:Z}

To calculate quantum corrections to solitons at one-loop order
in a well-defined manner,
we have to set up renormalization conditions and determine
counterterms to the parameters in the Lagrangian.
This is done by identifying the parameters and fields in
the Lagrangian as bare quantities $g_0$, $A_{(0)\mu}^a$, $a_{(0)\mu}^a$,
$\lambda_{(0)}$, \ldots,
and introducing renormalization constants according to
$g_0=Z_g g$,
$A_{(0)\mu}^a=\sqrt{Z_A} A_\mu^a$, $a_{(0)\mu}^a=\sqrt{Z_a} a_\mu^a$,
$\lambda_{(0)}=\sqrt{Z_\lambda} \lambda$, \ldots.
For this it is sufficient to consider
the spontaneously broken phase in the trivial background
with no monopoles. Choosing 
$\langle S_1^a \rangle = v \delta^a_3$
and expanding the Lagrangian with $R_\xi$ gauge
fixing term (\ref{Lgf}) and $\xi=1$ one finds
that all fields with color index $a=3$ are massless,
whereas $a=1,2$ have mass $m=gv$.

Because our gauge-fixing term is background covariant, we have
different vertices for bosonic background fields and bosonic
quantum fields. Considering firstly tadpole diagrams
with an external Higgs boson field,
we find that they vanish for an external background field $S_1$
as well as for an external quantum field $s_1$.
This is fortunate because in our model there is no counterterm
linear in $s_1$ or $S_1$ so that a nonvanishing tadpole divergence
could not have been removed by renormalization.

For determining wave-function and coupling constant renormalization
we shall only consider external background fields. The
background covariance of (\ref{Lgf})
then fixes the coupling constant renormalization
through wave-function
renormalization of the gauge boson background fields according 
to $Z_g=Z_A^{-1/2}$.
Using the two-point self-energy of the massless gauge boson
with external background fields
we require that ${Z_A}$ absorbs the entire one-loop
correction on mass-shell, which because of Lorentz invariance
is simply obtained at vanishing external momentum.
Because the massless gauge boson only couples to massive fields,
this does not involve infrared divergences,
and one finds in
dimensional regularization after a Wick rotation
\bea\label{g2ren}
Z_A&=&Z_g^{-2}=1+2(4-N)g^2I,
\nn
I&\equiv&\int {d^{4+\epsilon}k\0(2\pi)^{4+\epsilon}}
{-i\0 (k^2 + m^2)^2 }=
\int {d^{4+\epsilon}k_E\0(2\pi)^{4+\epsilon}}
{1\0 (k^2_E + m^2)^2 },
\eea
where $N=2$ or 4,
and $k_E$ are Euclidean 4-momenta. 
Note that in the case $N=4$ the coupling constant
and the (background) gauge field wave function do not renormalize,
whereas for $N=2$ there is infinite ultraviolet renormalization
with $I^{\rm div}=-{1\08\pi^2}{1\0\epsilon}$.

Renormalizing similarly the massless Higgs boson on-shell,
one finds $Z_S=Z_A$ in the gauge $\xi=1$, 
but 
\be\label{ZS}
Z_S=1+2(4-N+1-\xi)g^2I\ee
for $\xi\not=1$. So for $\xi\not=1$ there is a need for
infinite scalar wave function renormalization also in $N=4$.
(Divergent wave-function renormalizations in $N=4$
were previously obtained in Ref.~\cite{Kovacs:1999fx} 
for non-background-covariant gauges.)

The renormalization of the mass $m=gv$ of
the fields with SU(2) index $a=1,2$ is fixed by $Z_v=Z_S$ and
$m=Z_m m_0=Z_g Z_S^{1/2}g_0 v_0$ so that for $\xi=1$
the mass does not renormalize. However,
for general $\xi$ one has $Z_m=1+g^2(1-\xi)I$ as we have
verified by direct calculation
(contradicting a different assertion in Ref.~\cite{Imbimbo:1985mt}).

For the wave function renormalization of the fermions
we do not have to distinguish between quantum and background
fields. However, while the gauge fixing term (\ref{Lgf})
respects background gauge covariance, it is not susy invariant,
so the wave function renormalization constant $\sqrt{Z_\lambda}$
of the fermions is not related to $Z_A$ or $Z_S$, even when $\xi=1$.
A straightforward calculation of the one-loop
fermion self-energy, renormalized such that it vanishes
on-shell for the massless fermions, leads to
\be\label{Zpsi}
\lambda_0=\sqrt{Z_\lambda}\lambda,\qquad
Z_\lambda=1-2 (N+\xi-1) g^2 I
\ee
in $N$-extended super-Yang-Mills theory.
This is divergent even for $\xi=1$ and $N=4$.

In the following we shall restrict ourselves to the case $\xi=1$
as this is where we can employ index theorems to determine
spectral densities and thus perform explicit calculations
in the nontrivial (monopole) sector.

When considering composite operators such as the susy current
and its susy variations, one has to allow for composite-operator
renormalization. This is generally not simple multiplicative
renormalization, but involves matrices of renormalization constants
for a whole set of composite operators.

In Ref.~\cite{Hagiwara:1979pu} it has been shown that
the improved susy current $j_{\rm impr}^\mu=j^\mu+\Delta j_{\rm impr}^\mu$
is a finite operator and thus protected from renormalization.
However, in Ref.~\cite{Rebhan:2005yi} we have found the
somewhat surprising result that while in the case of $N=2$
both the unimproved and the improved current are finite
operators, not requiring renormalization beyond standard
wave function and coupling constant renormalization, the
situation is different in the $N=4$ theory.

By considering explicitly the renormalization of the
central charge operator $U$ in eq.\ (\ref{UUVV}), we have found that
for $N=4$ it requires nonmultiplicative renormalization
through improvement terms, similarly to what has
been observed in nonsupersymmetric theories
in the renormalization of the
energy-momentum tensor \cite{Callan:1970ze,Freedman:1974gs,Freedman:1974ze,Collins:1976vm,Brown:1980pq}.
However,
because the improved susy current does not renormalize,
it is only the improvement terms that acquire
composite-operator renormalization, and it turns out
that one can renormalize the latter multiplicatively.
In Ref.~\cite{Rebhan:2005yi} we have obtained
by explicit calculation 
\bea\label{ZDeltaU}
Z_{\Delta U_{\rm impr}}^{\rm{N=2}}&=&1,\nn
Z_{\Delta U_{\rm impr}}^{\rm{N=4}}&=&1+12g^2I,
\eea
where we have renormalized again with external massless
legs taken at vanishing momenta (which involves only massive
loop integrals). 
This result is in fact $\xi$-independent \cite{Rebhan:2005yi}.
Because the susy
violation that is implied by the gauge fixing term (\ref{Lgf})
and the corresponding ghost Lagrangian is only
in the form of BRS-exact terms,
the entire susy multiplet of
improvement terms has the same $Z$ factor,
\be
Z_{\Delta j^\mu_{\rm impr}}
=Z_{\Delta T^{\mu\nu}_{\rm impr}}=Z_{\Delta U_{\rm impr}}.
\ee

A consequence of this multiplicative renormalization of
improvement terms in the $N=4$ susy current is that
the unimproved susy current $j^\mu=j_{\rm impr}^\mu-\Delta j_{\rm impr}^\mu$
receives an {\em additive} renormalization with counterterm
\be\label{deltajunimpr}
\delta j^\mu=-(Z_{\Delta U}-1)\Delta j_{\rm impr}^\mu.
\ee

\section{Bulk and surface contributions to the
quantum mass of susy monopoles}

\label{sec:bulksurf}

In this section we decompose the space integral of the Hamiltonian
density for monopoles into surface terms and bulk terms. To
one-loop order, the latter
correspond to the usual sum over zero-point energies,
but the former lead to new effects which are not present in
lower-dimensional solitons and 
which we evaluate in section \ref{sec:surfterms}. We always begin
with the $N=2$ case, and then give the corresponding results
for the $N=4$ case.

In order to determine the energy-momentum tensor, we consider
the quantum action 
in curved space. For the gauge fixed theory and $\xi=1$ it reads,
for the $N=2$ model
\bea\label{curvedspaceaction}
S&=&\int d^4x \Bigl\{ -\4 \sqrt{-g} g^{MR} g^{NS} F_{MN}(A+a)F_{RS}(A+a)\nn
&&-\sqrt{-g} \5\lambda \Gamma^M D_M(A+a)\lambda
-\2{1\0\sqrt{-g}}[D_M(A)\sqrt{-g}g^{MN}a_N]^2\nn
&&-[D_M(A)b]\sqrt{-g}g^{MN}D_N(A+a)c\Bigr\}.
\eea
In the $N=4$ case $\lambda$ is Majorana and an extra factor
$\2$ has to be included in the fermionic term. 

The matrix $\Gamma^M$ is defined as the product of the constant matrices 
in (\ref{Gammas}) and a vielbein field, but
the covariant derivatives $D_M(A)$ in (\ref{curvedspaceaction})
do not depend on the gravitational field. 
The stress tensor is defined by
\be
T_{MN}=-2{\delta\0\delta g^{MN}}S
\ee
(and a similar formula with the vielbein field for spinors).
We thus obtain, for $N=2$ and on-shell (see also footnote \ref{fn6}),
\bea
T_{MN}&=&F_{MS}F_N{}^S-\4\eta_{MN} F_{RS}F^{RS} +
\2\5\lambda\Gamma_N\olrD_M\lambda \nn
&&-2 a_{(M} \hat D_{N)} \hat D_R a^R+
\eta_{MN}[a^S \hat D_S \hat D_R a^R+\2(\hat D_R a^R)^2]\nn
&&+2[\hat D_{(M}b]D_{N)}c-\eta_{MN}[\hat D^R b]D_R c,
\eea
where $F_{MN}$ and $D_M$ involve the full field $A+a$ and where
we have temporarily introduced the notation $\hat D_M=D_M(A)$
for the background covariant derivative.
For $N=4$ the fermionic part of the stress tensor is instead
$T_{MN}^{\rm ferm}=\2\5\lambda\Gamma_N D_M\lambda$.

For the Hamiltonian density $T_{00}$ the terms quadratic in quantum fields
are given by
\bea\label{T002}
T_{00}^{(2)}&=&[\2(F_{0{\3S}})^2+\4 F_{\3R\3S}^2]^{(2)}
-2 a_0 D_0 D_R a^R - a^S D_S D_R a^R -\2 (D_R a^R)^2\nn
&&+2(D_0b)(D_0c)+(D^Rb)(D_Rc)
+T_{00}^{(2){\rm ferm}}
\eea 
where the indices {\small$ \3R, \3S$} run over the nonzero values
of the indices $R, S$. We have dropped the hat on $D$ as from here on
the covariant derivatives involve only the background fields.
The fermionic contribution is
given by
\be
T_{00}^{(2){\rm ferm}}=
\cases{\2\5\lambda\Gamma_0 {\olrD}_0\lambda &
for $N=2$, \cr
\2\5\lambda\Gamma_0 D_0\lambda & for $N=4$. \cr}
\ee

The bosonic contributions have the same formal structure in $N=2$
and $N=4$, only the range of the indices is different.
The first two terms in
(\ref{T002}) are the classical contribution to $T_{00}^{(2)}$
from the bosons. Expanding
\be
F_{0{\3S}}(A+a)=F_{0{\3S}}(A)+D_0(A)a_{\3S}-D_{\3S}(A)a_0+g a_0\times a_{\3S}
\ee
we obtain
\bea
[\2(F_{0\3S})^2]^{(2)}&=&-\2a_{\3S}D_0^2a_{\3S}-\2a_0D_{\3S}^2a_0+a_{\3S}D_{\3S}D_0a_0
+2gF_{0\3S}a_0\times a_{\3S}\nn
&&+\2\6_0(a_{\3S}D_0a_{\3S})+\2\6_s(a_0 D_s a_0)-\6_0(a_{\3S}D_{\3S}a_0),
\eea
where we wrote all 
terms involving two derivatives 
in the form $aDDa$ (``bulk terms'') plus total derivatives,
and used $a_{\3S}D_0D_{\3S}a_0=a_{\3S}D_{\3S}D_0a_0+a_{\3S}(gF_{0\3S}\times
a_0)$. 
In the total derivatives we reduced the range of the indices {\small$\3R,\3S$} to
three-dimensional ones: $r,s=1,2,3$.
Similarly,
\bea
[\4 F_{\3R\3S}^2]^{(2)}&=&-\2a_{\3S}D_{\3R}^2a_{\3S}+\2a_{\3S}D_{\3S}D_{\3R}a_{\3R}
+gF_{\3R\3S}(a_{\3R}\times a_{\3S})\nn
&&+\2\6_r(a_{\3S}D_r a_{\3S})-\2\6_r(a_{\3S}D_{\3S}a_r).
\eea

Expanding accordingly the contributions from the gauge fixing term yields
\bea
T_{00}^{g.f.(2)}&=&-2a_0D_0(D_{\3S}a_{\3S}-D_0a_0)-\2 a_0D_0^2a_0 \nn
&&+a_{\3S}D_{\3S}D_0a_0-\2a_{\3R}D_{\3R}D_{\3S}a_{\3S} \nn
&&-\2\partial_{r}(a_{r}D_{\3S}a_{\3S})-\2\partial_0(a_0D_0a_0)+\partial_0(a_0D_{\3S}a_{\3S}) .
\eea
Using that
\be
a_{\3S}D_{\3S}D_0a_0-a_0D_0D_{\3S}a_{\3S}=
-\6_0(a_0D_{\3S}a_{\3S})+\6_s(a_sD_0a_0)
\ee
we find for the final bulk terms, adding also the contributions
from the fermions and the ghosts,\bea
T_{00}^{(2){\rm bulk}}&=&T_{00}^{(2){\rm ferm}}
-\2a_{\3S}(D_0^2+D_{\3R}^2)a_{\3S}+a_0(\8{3\02}D_0^2-\2D_{\3S}^2)a_0\nn
&&+2 gF_{0\3S}(a_0\times a_{\3S})+gF_{\3R\3S}(a_{\3R}\times a_{\3S})
\nn
&&-\2b D_0^2 c-\2 (D_0^2b)c-\2 bD_{\3S}^2 c-\2 (D_{\3S}^2 b)c\nn
&=&T_{00}^{(2){\rm ferm}}+a_0D_0^2a_0-a_{\3S}D_0^2a_{\3S}
+2gF_{0\3S}(a_0\times a_{\3S})
-bD_0^2 c-(D_0^2b)c.
\eea
In the last line we have used the linearized field equations, which
in the background-covariant $\xi=1$ gauge read
\bea\label{boslinfeq}
&D_0^2a_{\3R}=D_{\3S}^2a_{\3R}+2gF_{\3R\3S}\times a_{\3S}
-2gF_{\3R0}\times a_{0},\nn
&D_0^2a_{0}=D_{\3S}^2a_{0}+2gF_{0\3S}\times a_{\3S},\nn
&D_0^2(b,c)=D_{\3S}^2(b,c).
\eea

For the total derivative terms we get
\bea
T_{00}^{(2){\rm tot.deriv.}}&=&
\2\6_0 (a_{\3S} D_0 a_{\3S}) +\2\6_s(a_0 D_s a_0)-\6_0(a_{\3S} D_{\3S} a_0)\nn
&&+\2\6_r (a_{\3S}D_r a_{\3S})-\2\6_r(a_{\3S} D_{\3S} a_r)
-\2\6_r(a_r D_{\3S} a_{\3S})\nn
&&-\2\6_0(a_0D_0a_0)+\6_0(a_0D_{\3S}a_{\3S})-2\6_0(a_0D_{\3S}a_{\3S})
+2\6_r(a_rD_0a_0)\nn
&&+\2\6_0^2(bc)+\2\6_r^2(bc)
\eea
Since the expectation value of products of quantum fields
is time-independent in a soliton background, we can drop
total $\6_0$ derivatives. Using furthermore $a_0 D_{\3S} a_0
=\2\6_s a_0^2$ and $a_{\3S}D_r a_{\3S}=\2\6_r a_{\3S}^2$,
the result for the surface terms simplifies to
\be\label{T00surf}
T_{00}^{(2){\rm tot.deriv.}}=
\4\6_s^2 a_0^2+\4 \6_r^2 a_{\3S}^2+\2\6_r^2(bc)
-\2\6_r\6_s(a_r a_s)+2\6_r(a_r D_0 a_0).\quad
\ee
In the special case of a monopole (rather than a dyon) background,
the 
last term in (\ref{T00surf})
plays no role at the one-loop level,
because in a monopole background there is no propagator
between $a_r$ and $a_0$.

Let us now also specialize the
bulk contribution to a monopole background.
We then have $F_{0\3S}=0$ and $D_0=\6_0$.
Furthermore, $F_{\3R\3S}$ is nonzero only
when {\small$\3R,\3S$} are both from the set of spatial
indices 1,2,3 plus the index corresponding to
the Higgs field (6 for the $N=2$ case, one of $5,6,7$ for the $N=4$ case).
As in sect.~\ref{sec:fluct}
we denote indices from the latter set by $\9m,\9n$,
so that $a_{\9m}=(a_m,s)$.
The bosonic linearized field equations (\ref{boslinfeq})
then decompose into
\bea\label{boslineqam}
&D_0^2a_{\9m}=\6_0^2a_{\9m}=D_{\9n}^2a_{\9m}+2gF_{\9m\9n}\times a_{\9n},\nn
\label{boslineqrest}
&\6_0^2(a_0,p,b,c;q_I)=D_{\9s}^2(a_0,p,b,c;q_I).
\eea
Here $p$ is the pseudoscalar field, forming a
quartet with $a_0,b,c$. For 
the fluctuations of the extra four scalar and
pseudoscalar fields of the $N=4$ theory
we have introduced the notation $q_I$
with $I=1,\ldots,4$ for $N=4$.
This index $I$ has the same range but of course otherwise
nothing in common with the SU(4) index on the dimensionally
reduced Majorana spinors $\lambda^I$.

Explicitly, we have for the bulk contribution to the Hamiltonian density
in the $N=2$ case
\be\label{T002bulk}
T_{00}^{(2){\rm bulk,N=2}}=
-a_{\9m}\6_0^2a_{\9m}+a_0\6_0^2a_0 - p\6_0^2p -b\6_0^2c-(\6_0^2b)c+{i\02}\psi^\dagger
\olrp_0 \psi,
\ee
and in the $N=4$ case
\bea\label{T002bulk4}
T_{00}^{(2){\rm bulk,N=4}}&=&
-a_{\9m}\6_0^2a_{\9m}+a_0\6_0^2a_0 - p\6_0^2p 
-b\6_0^2c-(\6_0^2b)c\nn
&&-q_I\6_0^2q_I+{i\02} (\lambda^I)^T\6_0\lambda^I.
\eea

The evaluation of the expectation value of $T_{00}$ requires
regularization. Our method is to use dimensional regularization,
where we extend the range of the spatial indices
to $1,\ldots,3+\epsilon$,
with $0\le\epsilon\le1$
and the extra dimension being chosen in one of the directions
where the monopole background can be trivially embedded, i.e.\ one of
the directions not involving the gauge field component $A_I$ which
has been chosen to accommodate the scalar field of the monopole
(or the nonvanishing expectation value in the trivial sector).
For definiteness we shall indicate the extra dimension from
where we descend by continuous dimensional reduction by the index 5.

Inserting the mode expansions (\ref{eq:fqf}) 
and (\ref{eq:a0}) in the expectation value of
$T_{00}^{(2){\rm bulk}}$ and integrating
over space\footnote{Spatial 
integration with respect to the extra $\epsilon$ dimensions
would give simply a factor $L^\epsilon$ where $L$ is the extent
of the (trivial) extra dimension.}
one finds that it has the form of a sum
over zero-point energies
\bea\label{E0pt}
&&M^{(1)\rm bulk}=\int \langle T_{00}^{(2){\rm bulk}} \rangle d^3x=
\int d^3x \int {d^3k\,d^\epsilon\ell \0 (2\pi)^{3+\epsilon}}
{\omega\02}(\mathcal N_+|\chi^+_k|^2(x)+\mathcal N_-|\chi^-_k|^2(x))\nn
&&\qquad\qquad=\mathcal N
\int d^3x \int {d^3k\,d^\epsilon\ell \0 (2\pi)^{3+\epsilon}}
{\omega\02}(|\chi^+_k|^2(x)-|\chi^-_k|^2(x))
\eea
where we recall that $\omega=\sqrt{\omega_k^2+\ell^2}$.
The factor $\mathcal N=\mathcal N^+=-\mathcal N^-$ arises
for the two cases of $N=2$ and 4 as follows.
For $N=2$ we have
$\mathcal N^+=4-2=2$ coming from $a_{\9m}$ and $\psi_+$,
and $\mathcal N^-=1+1-1-1-2=-2$,
coming from $a_0,p,b,c$, and $\psi_-$.
For the $N=4$ theory there is
a complete cancellation, because then
$\mathcal N^+=4-4=0$
and $\mathcal N^-=1+1-1-1+4-4=0$, with the extra $+4$ in $\mathcal N^-$
coming from the extra scalars $q_I$.
The term with
$a_0$ in (\ref{T002bulk}) contributes the same way
as $p$ to the mass because the different sign in (\ref{T002bulk})
is taken care of by the negative sign in
$[a_0(k,\ell),a_0^\dagger(k,\ell)]=-\delta(k,k')\delta(\ell,\ell')$.
For the ghosts one gets an extra minus sign because the
propagator is $\langle c b\rangle$, not $\langle b c\rangle$,
and $b$ and $c$ anticommute.

We note that
in (\ref{E0pt}) any discrete modes can indeed be dropped --- zero modes
lead to scaleless integrals which vanish in dimensional regularization,
whereas any massive modes for bound states
cancel between fermions and bosons.
Note also that (\ref{E0pt}) is well-defined only because
of the difference $(|\chi^+_k|^2(x)-|\chi^-_k|^2(x))$.
Treating the two contributions separately
one could not easily integrate first
over $x$ and afterwards over momenta, as we shall do eventually.
However, in the combined expression we can interchange the order
of the integrations and use index theorem techniques to evaluate
the spectral density
\be
\label{drh}
\Delta\rho(k^2)=\int d^Dx (|\chi^+_k|^2(x)-|\chi^-_k|^2(x)),
\ee
with $D=3$ for the monopole model, but in the following we shall
also consider the analogous situation for lower-dimensional
solitons.
\section{Index theorems and susy spectral densities}

\label{sec:indexth}

As we have seen,
for susy solitons (and susy instantons as well) the linearized
field equations for the quantum fields display a universal and
simple pattern: one-half of the fermions have the same field
equations as some of the bosons, whereas the field equations of the other
half of the fermions coincides with those
of the rest of the bosons and these
differ only by an interaction term without
derivatives. (If gauge fields are present, one may choose a gauge
fixing term such that the field equations of the 
quartet of ghosts, antighosts, timelike gauge fields $a_0$, and
pseudoscalar field $p$ are all equal.) Moreover, every solution
of the first field equation with nonzero eigenvalue corresponds to
a solution of the second field equation with the same eigenvalue,
and vice-versa. Taking into account that for fermions the number
of degrees of freedom is half the number of field components,
one might conclude that for susy solitons and instantons the sum over
nonzero modes of all bosons and fermions cancels.\footnote{For the
quartet $(a_0,b,c,p)$ this cancellation does indeed occur because the
ghosts and antighosts contribute terms with a negative sign,
whereas for $a_0$ and $p$ the sign is positive. For $a_0$ this
comes about because of two sign changes: the stress tensor
has an extra minus sign, but also $[a_0,a_0^\dagger]$.}
However, this conclusion is false in general. On a compact space it would
be correct (with suitable boundary conditions), 
but on an open space the density of states of the
first and the second field operator may be, and is in some cases,
different. The continuous spectrum can be labeled by a vector $\vec k$
which corresponds to the distorted plane waves of the scattering
states. If these solutions of the field equations depend on time
through a factor $e^{-i\omega t}$, the eigenvalues of the
continuous spectrum of each field operator are equal to
$\omega^2=\vec k^2+m^2$, where 
$m$ is the mass of the quantum fields far away from the solitons. 
Let us denote the density of states
of the first and the second field operator by $\rho^+(k^2)$
and $\rho^-(k^2)$, respectively.
Then the sum of zero-point energies of the continuous spectrum
can be written as
\be
M^{(1)\rm bulk}
={\rm Tr}\,\2\hbar\omega
=\mathcal N\int \2\hbar\omega\left( \rho^+(\1k^2) - \rho^-(\1k^2) \right)
{d^D\1k\0(2\pi)^D}
\ee
in $D+1$ dimensional Minkowski spacetime, where $\mathcal N$ depends
on the number of components of the spinors.
The difference $\Delta\rho=\rho^+(\1k^2) - \rho^-(\1k^2)$ vanishes
in some cases such as the $N=2$ vortex, the instanton, 
and the $N=4$ monopole, but
not in other cases such as the minimally supersymmetric kink
and the $N=2$ monopole, where it gives rise to anomalous
contributions to the mass and central charge of these solitons.

In the case of the susy kink and the susy vortex, it is possible
to evaluate $\Delta\rho$ rather easily \cite{Goldhaber:2004kn}
by rewriting it into a surface
term after expressing $\chi^-_{\1k}(\1x)$ in terms
of $\chi^+_{\1k}(\1x)$ by using the Dirac operator and
by using the known asymptotic values of $\chi^+_{\1k}(\1x)$.
However, for the higher-dimensional cases, the mode expansions
become progressively more complicated. One would prefer to
deal with the fields themselves, instead of the mode functions.
Here another method brings relief,
revealing moreover an intriguing relationship
of a nonvanishing $\Delta\rho$ to anomalies.

The basic idea is to start with the spatial components
of the axial vector
current $j_\mu=\5\psi\gamma_5\gamma_\mu\psi$ where the two halves of the
components of $\psi$ have field equations corresponding to
the two sets under consideration. 
Taking the vacuum expectation value reduces the problem from $D+1$
dimensions to one in $D$ spatial dimensions.
The space-divergence of
its vacuum expectation value, $\6/\6x^i {\,\rm Tr\,} \gamma_5\gamma_i
\<\psi(x)\5\psi(y)\>$ can be written as a bulk integral involving $\Delta\rho$,
but at the same time one can evaluate the space integral of
$\6_i j_i$ explicitly, because in a perturbation expansion in
the inverse radius $r^{-1}$ only a few terms contribute.
Because the bulk integrals involving $\Delta\rho$ correspond to
Feynman graphs which are power-counting divergent, one
must regularize these integrals. In a series of papers, E.\
Weinberg has tackled this problem, using Pauli-Villars regularization
for vortices \cite{Weinberg:1981eu} and monopoles \cite{Weinberg:1979ma}.
For instantons, L.S.\ Brown et al.\ \cite{Brown:1977bj}
have given a simple algebraic argument that $\Delta\rho=0$.
However, as we shall argue below, the same argument applies to kinks
where we know already from a direct calculation that $\Delta\rho\not=0$.
In Ref.\ \cite{Weinberg:1979ma} Weinberg has given an argument
why $\Delta\rho=0$ for instantons: the leading (logarithmic)
divergence cancels due to the presence of the selfdual antisymmetric
't Hooft symbols. In the following we shall present a uniform
treatment of solitons and instantons, which covers kinks,
vortices, monopoles as well as 
instantons.
From the point of view of this paper, the relevance of this
analysis is that it yields the spectral density $\Delta\rho$ that is required
in the evaluation of the bulk contributions to the quantum
mass and central charge of susy monopoles.


The field operators for the fermionic modes in the soliton
background without extra dimensions can be written in all cases 
in the following {\it generic} form
(see (\ref{eq:dslash}) for the case of monopoles)
\be\label{sfDDbar}
\7{\mathsf D}\psi_++i\omega\psi_-=0,\qquad \7{\5{\mathsf D}}\psi_-+i\omega\psi_+=0.
\ee
Consider for example the $N=1$ susy kink in 1+1 dimensions,
or any solitonic model for that matter with a potential
$V(\varphi)=\2U^2(\varphi)$. It contains a real scalar
field $\varphi$ and a Majorana spinor\footnote{Note that in
(\ref{eq:f1}) we defined $\psi=\left(\psi_+ \atop i\psi_-\right)$
for the monopole.}
$\left(\psi_+ \atop \psi_-\right)$,
and for a suitable representation of the Dirac matrices,
namely $\gamma^0=-i\sigma^2=\pmatrix{0&-1\cr1&0}$, 
$\gamma^1=\sigma^3=\pmatrix{1&0\cr0&-1}$, we find
for the Dirac equation
\be
\7{\mathsf D}\psi_+\equiv(\6_x+U')\psi_+=-i\omega\psi_-,\qquad 
\7{\5{\mathsf D}}\psi_-\equiv(\6_x-U')\psi_-=-i\omega\psi_+=0.
\ee
Iterating the Dirac equation 
and using the BPS equation $\6_x\varphi_K+U(\varphi_K)=0$
one finds
\bea
&&-\7{\5{\mathsf D}}\7{\mathsf D}\psi_+=(-\6_x^2+U'U'+UU'')\psi_+=\omega^2\psi_+,\nn
&&-\7{\mathsf D}\7{\5{\mathsf D}}\psi_-=(-\6_x^2+U'U'-UU'')\psi_-=\omega^2\psi_-,
\eea
where $U'$ and $U''$ are to be
evaluated for the kink background $\varphi_{\rm K}$.
Clearly, $\psi_+$ and $\eta=\varphi-\varphi_{\rm K}$ have the same
linearized field equation.

The field equations for $\psi_+$ and $\psi_-$ 
in the kink background can be written
in $2\times2$ matrix form by introducing an operator $\7{\mathcal D}$
\be\label{curlyDslashsusykink}
\7{\mathcal D}\left(\psi_+ \atop \psi_-\right)
=\pmatrix{0&\7{\5{\mathsf D}}\cr \7{\mathsf D} & 0}
\left(\psi_+ \atop \psi_-\right)
=\pmatrix{0&\6_x-U' \cr \6_x+U' & 0}
\left(\psi_+ \atop \psi_-\right)
\ee
where
\be\label{curlyDslashsusykink2}
\7{\mathcal D}=\sigma^j \mathcal D_j\qquad\mbox{with}\quad 
\mathcal D_1=\6_x,\;\mathcal D_2=-iU'.
\ee
The operator $\7{{\mathcal D}}$ is anti-hermitian,
$\7{\mathcal D}^\dagger=-\7{{\mathcal D}}$. Note that we
are considering the operators $\7{\mathsf D}$ and $\7{\5{\mathsf D}}$ as acting in
a one-dimensional space with coordinate $x$, although the spinor
space is two-dimensional.\footnote{%
For the susy kink, $\psi_+$ and $\psi_-$ are just the
components of the spinor field, and we shall have the same situation
in the case of the monopole, but for the vortex $\psi_+$ and $\psi_-$
are linear combinations of the matter and gauge fermions of this model
\cite{Rebhan:2003bu}.}

Index theorems are primarily used to compute the number of zero
modes (normalizable solutions to the linearized fluctuation equations
which are time-independent). For this purpose one considers
\be
\mathcal J(M^2)={\,\rm Tr\,}\left(
{M^2\0-\7{\5{\mathsf D}}\7{\mathsf D}+M^2}-{M^2\0-\7{\mathsf D}\7{\5{\mathsf D}}+M^2}
\right)
\ee
and defines the index of $\7{\mathsf D}$ as $\mathcal I=\lim_{M^2\to0} \mathcal J(M^2)$.
In the limit $M^2\to0$, only zero modes can contribute\footnote{Provided
the continuum spectrum is not too singular
in the infrared, but this requirement
is met in the cases we shall consider \cite{Weinberg:1979ma}.}
so that 
$\mathcal I$
gives the difference of the number of zero modes
of $\7{\5{\mathsf D}}\7{\mathsf D}$ and $\7{\mathsf D}\7{\5{\mathsf D}}$. In all models we consider, only
$\7{\5{\mathsf D}}\7{\mathsf D}$ has zero modes. E.g., for the kink with $U=(\lambda/2)^{1/2}
(\varphi^2-m^2/2\lambda)$, the potential $U'U'+UU''=(\2U^2)''=
3\lambda\varphi^2-m^2/2$
in $\7{\5{\mathsf D}}\7{\mathsf D}$ has a zero mode corresponding to the position of the
kink, but in $\7{\mathsf D}\7{\5{\mathsf D}}$ one finds 
the positive definite potential
$U'U'-UU''=\lambda\varphi^2+m^2/2$ without zero modes.
Both potentials tend to $m^2$ as $|x|\to\infty$,
which identifies $m^2$ as the asymptotic mass in terms of
which $\omega^2=k^2+m^2$.

Since, with the exception of zero modes, all eigenvalues
of $\7{\5{\mathsf D}}\7{\mathsf D}$ and $\7{\mathsf D}\7{\5{\mathsf D}}$ are the same,
the quantity $\mathcal J(M^2)$ is directly related
to the difference of the spectral densities of the continuum modes
\be\label{JDeltarho}
\mathcal J(M^2)-\mathcal J(0) \equiv \mathcal J_{\rm cont.}(M^2) =
\int{d^D\1k\0(2\pi)^D}
{M^2\0\omega^2+M^2}
\Delta\rho(\1k^2)
.\ee
One can rewrite $\mathcal J(M^2)$ as
a trace over a spinor space which is twice as large
\be
\mathcal J(M^2) = {\,\rm Tr\,}\left({M^2\0-\7{\mathcal D}^2+M^2}\Gamma_5\right)
\ee
where $\Gamma_5=\pmatrix{\mathbf1&0\cr0&-\mathbf1}$ is the chirality operator,
and $\7{\mathcal D}=\Gamma^j \mathcal D_j$ defines the Dirac
matrices $\Gamma^j$.
To denote the generic case we use capital $\Gamma$,
which should not be confused with the higher-dimen\-sional gamma matrices
in (\ref{eq:6dgam}) and (\ref{Gammas}).
[For the example of the
kink, $\Gamma_5=\sigma^3$ and $\Gamma^j=\sigma^j$,
see (\ref{curlyDslashsusykink2}). 
Note that we first decomposed the Dirac equation
into chiral parts, and then re-assembled these parts into
nonchiral expressions using another representation for the Dirac
matrices in (\ref{curlyDslashsusykink2}). 
Also this is a general feature of the procedure.]
To actually evaluate the continuum contribution to $\mathcal J(M^2)$,
one introduces plane waves as bras and kets with 
$\<\1k|\1k'\>=\delta^D(\1k-\1k')$, 
\be
\mathcal J(M^2) = {\,\rm tr\,}
\int d^Dx\int d^D k\int d^Dk'
\<\1x|\1k\>\<\1k|{M^2\0-\7{\mathcal D}^2+M^2}\Gamma_5|\1k'\>\<\1k'|\1x\>,
\ee
where the trace tr is over spinor and group indices only.
Then one pulls the ket plane wave $e^{-i\1k'\cdot\1x}/(2\pi)^{D/2}$
to the left, after which the derivatives $\6_j$ in $\7{\mathcal D}$
are replaced by $\6_j-ik'_j$, and $\<k|k'\>=\delta^D(k-k')$
sets $k'=k$.
The denominator can then be expanded as follows
\be
(\1k^2+M^2)+(2i\1k{\6\0\6\1x}-\mathcal D_j^2-\2\Gamma^j\Gamma^k \mathcal F_{jk})
\ee
where $[\mathcal D_j,\mathcal D_k]=\mathcal F_{jk}$. (For example, for the kink one obtains
\be\label{susykinkdenom}
(k^2+M^2)+(2ik{\6\0\6x}-\6_x^2+U'U'+i\sigma^1\sigma^2\6_x U')
\ee
where $\6_xU'=-U''U$ due to the BPS equation $\6_x\varphi_K+U(\varphi_K)=0$.)

At this point one encounters a technical problem which has a deep
theoretical meaning. In order to explicitly evaluate $\mathcal J(M^2\to0)$
one first needs to know $\mathcal J(M^2)$ itself. If (as is the case
in some models, but not in those of particular interest to us)
$\mathcal J(M^2)$ is $M^2$-independent, one can compute $\mathcal I$
by taking instead the limit $M^2\to\infty$
in $\mathcal J(M^2)$. In that case one
can expand the denominator in
\be
\mathcal J(M^2)=\int d^Dx\,\int {d^Dk\0(2\pi)^D}
{\rm tr}{M^2\Gamma_5\0(\1k^2+M^2)+(2i\1k{\6\0\6\1x}-
\mathcal D_j^2+\2\Gamma^j\Gamma^k \mathcal F_{jk})})
\ee
around $\1k^2+M^2$, and only a few terms contribute for $M^2\to\infty$.
This is the calculation first performed by Fujikawa \cite{Fujikawa:1983bg} 
to compute the chiral anomaly.

However, in some cases $\mathcal J(M^2)$ depends on $M^2$. In these
cases $\Delta\rho$ is nonvanishing. To still be able to compute $\mathcal J$,
one may try to relate $\mathcal J(M^2)$
to a surface integral, because then instead of doing perturbation
theory for $M^2\to\infty$ one can hope to do perturbation theory
for $r\to\infty$. At this point the axial anomaly comes to the rescue.
Recalling that for massive fermions the axial current
$j_5^\mu=\5\psi\Gamma_5\Gamma^\mu\psi$ satisfies
\be\label{axcurranomaly}
\6_\mu j^\mu_5=-2M\5\psi\Gamma_5\psi+{\rm anomaly}
\ee
and writing the fermion propagator as $\langle \psi(x)\5\psi(y)\rangle=
{-i\0\7{\mathcal D}+M}\delta(x-y)$, one finds for ${i\02}$
times the first term
on the right-hand side, using
$\tr \5\psi\Gamma_5\psi=-\tr \Gamma_5\psi\5\psi$,
\bea\label{JM2eq}
iM \int d^Dx \lim_{y\to x}
{\,\rm tr\,}\Gamma_5\langle \psi(x)\5\psi(y)\rangle&=&
M {\,\rm Tr\,}\Gamma_5{1\0\7{\mathcal D}+M} \nn
&=& {\,\rm Tr\,}\Gamma_5{M^2\0-\7{\mathcal D}^2+M^2}=\mathcal J(M^2).
\eea
We have then indeed an expression for $\mathcal J(M^2)$ as a total divergence
plus an anomaly term.

Once again one is confronted with a new subtlety. When there is an anomaly,
one should be careful and specify the regularization procedure.
We use Pauli-Villars regularization
with regulator mass $\mu$, which is the
most convenient for the technical step of extracting $\Delta\rho$.
When the latter is used in the evaluation of the quantum corrections
to the mass and central charge of the soliton, we shall
switch to dimensional regularization and define our renormalization
scheme in dimensional regularization. In the calculation
of $\Delta\rho$ we do not have to renormalize yet.

We are thus led to consider the following expression \cite{Weinberg:1981eu}
\be
J_i(x,y;M,\mu)=
{\,\rm Tr\,} \left(\langle x| \Gamma_5\Gamma_i {1\0\7{\mathcal D}+M}|y\rangle
-\langle x| \Gamma_5\Gamma_i {1\0\7{\mathcal D}+\mu}|y\rangle \right)
\ee
This is the current we consider from now on; the fermion fields
$\psi$ and $\5\psi$ themselves, in terms of which the axial current
was defined in (\ref{axcurranomaly}) will no longer be used. Note
that the index $i$ only takes on spacelike values. Hence, the whole
analysis is in Euclidean space.

In all cases considered below we construct an operator $\7{\mathcal D}$
from operators $\7{\mathsf D}$ and $\7{\5{\mathsf D}}$ as
in (\ref{curlyDslashsusykink}). These $\7{\mathcal D}$
consist of terms with ordinary derivatives $\Gamma^i\6_i$ and terms $K$
without these derivatives. From the off-diagonal structure of
$\7{\mathcal D}$ it follows that $\Gamma^i$ and $K$ anticommutate
with the chirality operator $\Gamma_5$.
The coefficients $\Gamma^i$ of $\6_i$ satisfy a Clifford algebra
\bea\label{curlyDslash}
&&\7{\mathcal D}=\Gamma^i \mathcal D_i\equiv \G^i\6_i+K,\nn
&&\{\Gamma^i,\Gamma^j\}=2\delta^{ij},\quad
\{\Gamma^i,\Gamma_5\}=0,\quad
\{K,\Gamma_5\}=0,\quad
\Gamma_5=\pmatrix{\mathbf1&0\cr0&-\mathbf1}.\qquad
\eea
We need an identity for the Green function
of $\7{\mathcal D}+M$;
this identity appears in Ref.~\cite{Weinberg:1981eu}, and we
only enumerate here the steps one needs to derive it. One starts
from
\bea
&&(\7{\mathcal D}_y+M)[\7{\mathcal D}_y+M]^{-1}\delta^D(x-y)
=\delta^D(x-y)\nn
&&[\7{\mathcal D}_y+M]^{-1}(\7{\mathcal D}_y+M)\delta^D(x-y)=\delta^D(x-y).
\eea
Then one replaces $(\7{\mathcal D}_y+M)\delta^D(x-y)$ in the second
identity by $\delta^D(x-y)(-\olp_{x^i}\Gamma^i+K(x)+M)$, replaces the
$\delta^D(x-y)$ in both relations by $\<x|y\>$ and pulls the bra $\<x|$
in the second relation
to the left past $[\7{\mathcal D}_y+M]^{-1}$. Next one multiplies
both relations by $\Gamma_5$ from the left, uses
(\ref{curlyDslash}) to move $\Gamma_5$ past $\7{\mathcal D}_y$
in the first relation, and takes the spinor trace. This yields
\bea
&&{1\02}\left({\6\0\6x^i}+{\6\0\6y^i}\right)J^i(x,y;M,\mu)\nn
&=&- M \tr \<x|\Gamma_5[\7{\mathcal D}_y+M]^{-1}|y\>
+\mu \tr \<x|\Gamma_5[\7{\mathcal D}_y+\mu]^{-1}|y\>\nn
&&-\2\tr \left[
(K(x)-K(y))\Gamma_5\<x|[\7{\mathcal D}_y+M]^{-1}-[\7{\mathcal D}_y+\mu]^{-1}|
y\> \right].
\eea
In the regulated identity we let $y$ tend to $x$, in which case the
last term vanishes. Through (\ref{JM2eq}) we finally find the
desired identity which expresses $\mathcal J(M^2)$ and thus $\Delta\rho$
in terms of a surface integral and the axial anomaly,
\be\label{Jsurfint}
-\2\int d^Dx{\6\0\6x^i}J^i(x,x;M,\mu)=\mathcal J(M^2)-\mathcal J(\mu^2).
\ee
We shall evaluate $\mathcal J(\mu^2)$ for large $\mu^2$ and
$\int d^Dx{\6\0\6x^i}J^i$ for large $|\1x|$.
In an odd number of dimensions there is no axial anomaly,
and we shall indeed find that for the kink and the monopole,
where there is an odd number of spatial dimensions, $\mathcal J(\mu^2)$
vanishes for $\mu^2\to\infty$. Any $M$-dependence of $\mathcal J$
is then due to the $M$-dependence of the surface term.

To evaluate the anomaly $\mathcal J(\mu^2\to\infty)$, we insert plane waves
as explained before, to obtain
\bea\label{Jmu2}
&&\mathcal J(\mu^2)=\int d^Dx \int {d^Dk\0(2\pi)^D}
\tr \Gamma_5 {\mu^2\0(k^2+\mu^2)-L}\nn
&&L=-2ik_j\6_j+\6_j^2+(\Gamma^iK+K\Gamma^i)(\6_i-ik_i)
+\Gamma^i(\6_iK)+K^2.
\eea
Expanding the denominator in terms of $L$, only a few terms
contribute for $\mu^2\to\infty$, and of these only
one term survives after taking the trace over spinor indices.

To evaluate the surface integral, we use that in all cases considered
the potential $K^2$ tends to a constant $m^2$ for $|\1x|\to\infty$.
We write then
\bea\label{Jixx}
J^i(x,x;M,\mu)&=&\int {d^Dk\0(2\pi)^D} \biggl\{ \tr
\Gamma_5 \Gamma^i (-\7{\mathcal D}+i\7k)[k^2+M^2+m^2-\ell]^{-1}\nn
&&-\tr\Gamma_5 \Gamma^i (-\7{\mathcal D})[k^2+\mu^2+m^2-\ell]^{-1} 
\biggr\}
\eea
with $\ell=L-m^2\equiv -2ik_j\6_j+\6_j^2+(\Gamma^iK+K\Gamma^i)(\6_i-ik_i)
+C$.
The matrix-valued functions $\Gamma^iK+K\Gamma^i$ and $C$ tend
to zero for large $\1x$ at least as fast as $|\1x|^{-1}$, and one
can evaluate the surface integral by expanding in terms of $\ell$.

\subsection{The susy kink}

For the susy kink the operator
$\7{\mathcal D}$ was given in (\ref{curlyDslashsusykink2}).
Using (\ref{susykinkdenom}) we obtain for the anomaly
\be
\mathcal J(\infty)=
\lim_{\mu^2\to\infty}
\int dx \int {dk\02\pi}
\tr
\sigma_3\mu^2[(k^2+\mu^2)+(2ik{\6\0\6x}-\6_x^2+U'U'-\sigma_3\6_x U')]^{-1}
\ee
which vanishes as expected.

For the surface integral we obtain
\bea
&&-\2\int dx {\6\0\6x}J(x,x;M,\infty)
=-\2\int dx {\6\0\6x} \int {dk\02\pi}
\tr{\sigma_3\sigma_1(-\sigma_1\6_x+i\sigma_2U')\0k^2+M^2+m^2}\nn
&&=U'\Big|_{x=-\infty}^{x=\infty}\int{dk\02\pi}{1\0k^2+M^2+m^2}
={m\0\sqrt{m^2+M^2}}
\eea
The index is given by $\mathcal I=\mathcal J(0)=1$,
and the corresponding zero mode is the zero mode for translations.

The spectral density $\Delta\rho$ is determined by (\ref{JDeltarho}),
\be\label{JDeltarhokink}
\mathcal J(M^2)-\mathcal J(0) = {m\0\sqrt{m^2+M^2}}-1 =
\int{dk\02\pi}
{M^2\0k^2+m^2+M^2}
\Delta\rho(k^2)
.\ee
One can solve this integral equation by a Laplace transform, and the
result is \cite{Kaul:1983yt,Imbimbo:1984nq,Callias:1977kg}
\be
\Delta\rho(k^2)={-2m\0k^2+m^2}.
\ee

Thus for the susy kink $\mathcal J(M^2)$ 
is $M$-dependent and
hence $\Delta\rho$ nontrivial.
This shows that the argument given in Ref.~\cite{Brown:1977bj}
which uses cyclicity of the trace
to prove the $M$-independence of $\mathcal J(M^2)$ for
instantons is inapplicable, because the kink provides a counter example.
Because in the kink there are no long-range massless fields,
also the explanation given in the appendix of
Ref.\ \cite{Weinberg:1979ma} of why the argument of Ref.~\cite{Brown:1977bj}
does not apply to the case of monopoles is incomplete.

\subsection{The $N=2$ susy vortex}

The $N=2$ vortex model in 2+1 dimensions
\cite{Schmidt:1992cu,Edelstein:1994bb,Lee:1995pm,Vassilevich:2003xk,Rebhan:2003bu}
contains one abelian gauge field, one complex scalar $\phi$, another
real scalar and
one two-component complex gaugino and matter fermion.
One can introduce linear combinations $U$ and $V$ of the
gaugino and matter fermion,
which can be viewed as the chiral parts of a complex four-component
spinor 
\be
\7{\mathsf D}U=-i\omega V,\quad \7{\5{\mathsf D}}V=-i\omega U
\ee
with
\be\label{sfDvortex}
\7{\mathsf D}=
\left(\begin{array}{cc}
D_+^{\mathrm V} & -i\sqrt2 e \pv \\
i\sqrt2 e \pv^*  & \6_-
\end{array}\right), \quad
\7{\5{\mathsf D}}=
\left(\begin{array}{cc}
D_-^{\mathrm V} & i\sqrt2 e \pv \\
-i\sqrt2 e \pv^*  & \6_+
\end{array}\right),
\ee
where $D_\pm=\6_\pm-ieA_\pm$, $\6_\pm=\6_1\pm i\6_2$, $A_\pm=A_1\pm iA_2$.
Quantities with sub- or superscript V refer to the
background fields of the vortex solution
\cite{Abrikosov:1957sx,Nielsen:1973cs,deVega:1976mi,Taubes:1980tm}, 
$A_j^{\mathrm V}(\1x)$
and $\pv(\1x)$ with $j=1,2$. Since $\7{\mathsf D}^\dagger=-\7{\5{\mathsf D}}$
we introduce a 4$\times$4 matrix-valued antihermitian operator
$\7{\mathcal D}$
\be\label{curlyDslashvortex}
\7{\mathcal D}
=\pmatrix{0&\7{\5{\mathsf D}}\cr \7{\mathsf D} & 0}
\equiv \Gamma^i\6_i+K.
\ee
The operator $\7{\mathcal D}\7{\mathcal D}$ contains the
operator $\7{\5{\mathsf D}}\7{\mathsf D}$ and
$\7{\mathsf D}\7{\5{\mathsf D}}$ along the diagonal, respectively, where

\be\label{DD1v}
\7{\5{\mathsf D}}\7{\mathsf D}
=\left(
\begin{array}{cc}
(D_k^{\mathrm V})^2-e^2(3|\pv|^2-v^2)  &  -i\sqrt2 e (D_- \pv) \\
i\sqrt2 e (D_- \pv)^* & \6_k^2-2 e^2 |\pv|^2
\end{array}
\right),
\ee
with $k=1,2$ and
\be\label{DD2v}
\7{\mathsf D}\7{\5{\mathsf D}}=\left(
\begin{array}{cc}
(D_k^{\mathrm V})^2-e^2(|\pv|^2+v^2)  &  0 \\
0 & \6_k^2-2 e^2 |\pv|^2
\end{array}
\right).
\ee
Only the operator (\ref{DD1v}) has zero modes, and this is also
the operator which governs the fluctuations of the
(complex) doublet $(\eta,a_+/\sqrt2)$, where $\eta=\phi-\pv$
and $a_+=A_+-A_+^{\rm V}$.
For large $|\1x|$, the operator $\7{\mathcal D}\7{\mathcal D}$
approaches the diagonal matrix $\{(D_k^{\mathrm V})^2-m^2,
\6_k^2-m^2,(D_k^{\mathrm V})^2-m^2,
\6_k^2-m^2\}$, $m^2=2e^2v^2$, exponentially fast.

We are now ready to compute the anomaly and the surface term for the vortex.
There is an anomaly in two dimensions, and it is given by
\be\label{eq111}
\mathcal J(\mu^2)=\int {d^2k\0(2\pi)^2}
{\mu^2\0(k^2+\mu^2)^2}\int d^2x\tr (\Gamma_5\Gamma^i\6_iK),
\ee
because $\tr \Gamma_5 K^2=-\tr K\Gamma_5 K=0$.
Since $\Gamma_5$ is diagonal, we only need the diagonal
entries of $\Gamma^i\6_iK$, which are easily read off from
(\ref{curlyDslashvortex}) and (\ref{sfDvortex}), and are given
by $(-ie\6_-A_+^{\rm V},0,-ie\6_+A_-^{\rm V},0)$.
This leads to
\be\label{eq112}
\int d^2x\, ie(\6_+A_-^{\rm V}-\6_-A_+^{\rm V})=
\int d^2x\, 2eF_{12}=4\pi n,
\ee
where $n$ is the winding number of the vortex background.
Using $(2\pi)^{-2}\int d^2k\,\break (k^2+\mu^2)^{-2}=(4\pi\mu^2)^{-1}$,
we find $\mathcal J(\mu^2)=n$.

The $M$-dependent part of the surface term is given by
\bea
-\2\int d^2x \6_i J^i&=&
\int d^2x \,\6_i \int{d^2k\0(2\pi)^2}\biggl[
(k^2+M^2+m^2)^{-1} \tr \Gamma_5 \Gamma^i(-K)\nn&&
+(k^2+M^2+m^2)^{-2} \tr \Gamma_5 \Gamma^i(-\Gamma^j\6_j-K)C \biggr].
\eea
The last term does not contribute since all entries of $C$
tend to zero exponentially fast as $r\to\infty$ 
\cite{Nielsen:1973cs,deVega:1976mi,Taubes:1980tm}. 
The first term also vanishes: according to Stokes' theorem
$\int d^2x \6_i v^i=
\int d^2x g^{-1/2}\6_i g^{ij}g^{1/2}v_j=\int d\theta x^i v_i$,
but the trace $\tr \Gamma_5 x^i \Gamma_i K$ is proportional
to $x^+ A_-^{\rm V}-x^- A_+^{\rm V}$
(replace $\6_i$ by $x^i$ in the step from (\ref{eq111}) to (\ref{eq112})), 
and this vanishes (see (\ref{asympt})).

The final result for $\mathcal J$ is therefore
$\mathcal J(M^2)=n$ independently of $M^2$. As we have discussed
before, this implies that $\Delta\rho$ vanishes for the vortex.
For the index we evidently have $\mathcal I=n$. Because the operator
(\ref{DD1v}) is the fluctuation operator of $(\eta,a_+/\sqrt2)$,
this implies that there are $n$ independent complex zero modes
for $\psi^+$,
or $2n$ real zero modes for $({\rm Re}\eta,{\rm Im}\eta,a_1,a_2)$
\cite{Rebhan:2004vn}.
These correspond to the positions in the $x$-$y$ plane of $n$
simple vortices . (The zero modes associated with rotations
of the vortex solutions are not new zero modes, but linear
combinations of the translational ones; one can prove this
in the same way as for instantons \cite{Belitsky:2000ws}).

\subsection{The $N=2$ susy monopole}

For calculating $\Delta\rho$ for the $N=2$ monopole,
regularization is not needed,\mpar{Regularization would be
needed for Brown's cyclicity trick!}
but we keep the Pauli-Villars contributions as
a check. The fermionic fluctuation equations have again the form
(\ref{sfDDbar}), with $\7{\sf D}$ and $\5{\7{\sf D}}$ given 
in (\ref{eq:dslash}), and we construct again the 4$\times$4
matrix $\7{\mathcal D}$, whose decomposition $\7{\mathcal D}=\G^i\6_i+K$
defines $K$ and $\Gamma^i$, satisfying (\ref{curlyDslash}).
There are now two new features: (i) because the square of the scalar
triplet $S^2$ tends to $v^2$ for large $r$ as slowly as
$S^2=v^2(1-2(mr)^{-1}+\ldots)$ (see (\ref{asympt})),
there will be surface contributions
from these scalars; (ii) the propagator $(-\7{\mathcal D}\7{\mathcal D}
+M^2)^{-1}$ must be decomposed into two separate isospin sectors.

The anomaly term is given by (\ref{Jmu2}) in the limit $\mu^2\to\infty$,
but this should vanish, because there are no chiral anomalies
in odd (three) dimensions. (Anomalies will however appear when
we later use the spectral densities to evaluate the quantum
corrections to mass and central charge of the monopole.)

The surface term is given by (\ref{Jsurfint}) with (\ref{Jixx}),
which reads more explicitly as follows
\bea
&&-\2\int d^3x\,\6_iJ^i(x,x;M,\mu)=-\2\oint d\Omega\,r^2
\int{d^3k\0(2\pi)^3} \tr \Bigl\{ \hat x^i \Gamma_5 \Gamma^i
(i\Gamma^j k_j-\7{\mathcal D})^{ab}\nn
&&[(k^2+M^2+m^2-\ell_+)^{-1}_{bc}\left(\delta^{ca}-\hat x^c \hat x^a\right)
+(k^2+M^2+m^2-\ell_+)^{-1}_{bc}\hat x^c \hat x^a]\Bigr\}_{r\to\infty}\nn
&&\qquad - \{ M^2\to\mu^2 \}
\eea
where the trace is over spinor indices, and
\bea
\ell_-^{ab}&=&-2ik_j\6_j \delta^{ab}+(D_j^2)^{ab}
+(g^2S^2-m^2)\delta^{ab}\nn
\ell_+^{ab}&=&\ell_-^{ab}+\2g \Gamma^{\9m\9n}F_{\9m\9n}^{ab}
\eea
with $F_{\9m\9n}^{ab}=g\epsilon^{acb}F^c_{\9m\9n}$
and
$
F_{\9m\9n}=\pmatrix{ F_{mn} & D_m S \cr -D_n S & 0 }.
$
The Dirac matrices which appear in these expressions are given by
\be
\Gamma^{\9m}=\pmatrix{0 & \bar \sigma^{\9m} \cr \sigma^{\9m} & 0},
\quad \Gamma^1\Gamma^2\Gamma^3\Gamma^4=-\Gamma^5,\quad
\Gamma^5=\pmatrix{\mathbf1&0\cr0&-\mathbf1}
\ee
and we have rewritten the term $\2\bar\sigma^{\9m\9n}F_{\9m\9n}$
from (\ref{eq:d2}) as the 4$\times$4 matrix $\2\Gamma^{\9m\9n}F_{\9m\9n}$
in order to deal with only one kind of Dirac matrices. The
curvature $F_{mn}$ falls off as $1/r^2$, and coming from
the expansion of $\ell_+$ is the only term which contributes.
Inserting the asymptotic values of the monopole background field (\ref{asympt})
we find for the $M^2$-dependent part
\bea\label{JM2mon}
\mathcal J(M^2)&=&\lim_{r\to\infty}\int d\Omega \,r^2
\int {d^3k\0(2\pi)^3}{1\0(k^2+M^2+m^2)^2}\nn
&&\times
\tr \2\Gamma_5 \hat x^i \Gamma^i \Gamma^4 i S^{ab}
(\2g\Gamma^{jk}F_{jk})^{ba}={2m\0\sqrt{M^2+m^2}}.
\eea
(The momentum integral yields $(8\pi\sqrt{M^2+m^2})^{-1}$
and the trace tends to $4m/r^2$ as $r\to\infty$.)
The contribution due to the Pauli-Villars regulator is proportional
to $(\mu^2+m^2)^{-1/2}$ and indeed vanishes for $\mu^2\to\infty$.

The result for the index is $\mathcal I=\mathcal J(0)=2$, corresponding
to 2 complex zero modes for the fermions, and 4 real zero modes for
a monopole with winding number
unity: 3 zero modes for its position, and one for gauge
orientations with respect to the unbroken U(1).
(The latter is given by $\delta A_i=(D_iS)\alpha$ and $\delta S=0$
with $\alpha$ the modulus \cite{Weinberg:1979ma}).

Finally we extract the result for $\Delta\rho$ from (\ref{JM2mon}).
According to (\ref{JDeltarho}) we must invert
\be
{2m\0\sqrt{M^2+m^2}}-2=\int {d^3k\0(2\pi)^3} {M^2\0k^2+m^2+M^2}
\Delta\rho_{\rm mon}(k^2).
\ee
The solution is
\be\label{drh2}
\Delta\rho_{\rm mon}(k^2)={-4\pi m\0k^2(k^2+m^2)}.
\ee


\section{Results for the quantum monopole mass}

In sect.~\ref{sec:bulksurf} we have shown that the contributions
to the mass of the $N=2$ and $N=4$ monopole consist of bulk
terms which correspond to the familiar sum over zero-point energies
of the bosonic and fermionic quantum fluctuations, and surface terms
whose presence seems at first sight unusual and puzzling. 
As we shall see later in this section, the latter have the
potential to contribute nontrivially to the one-loop quantum mass
of monopoles, in contrast to lower-dimensional solitons.
First, however, we shall use the results of the previous section
to evaluate the quantum corrections to the monopole mass
from the sum over zero-point energies \cite{Rebhan:2004vn}.

\subsection{Bulk contributions}

In (\ref{E0pt}) we have found that in dimensional regularization
the contributions from $T_{00}^{(2){\rm bulk}}$ have the form
\be\label{E0ptmon}
M^{(1)\rm bulk}=\int \langle T_{00}^{(2){\rm bulk}} \rangle d^3x=\mathcal N
\int d^3x \int {d^3k\,d^\epsilon\ell \0 (2\pi)^{3+\epsilon}}
{\sqrt{k^2+\ell^2+m^2}\02}\Delta\rho(k)
\ee
where $\mathcal N=4-N$ for the $N=2$ and $N=4$ theory,
and $\Delta\rho$ given by (\ref{drh2}).

Clearly $M^{(1)\rm bulk}$ is logarithmically divergent.
Carrying out the $\ell$-integration we obtain\footnote{Use
$\int d^np/(p^2+M^2)^\alpha=\pi^{n/2}(M^2)^{n/2-\alpha}\Gamma(\alpha-n/2)
/\Gamma(\alpha)$.}
\be\label{Ebulkres}
M^{(1)\rm bulk}=-(4-N)
{m\0\pi} {\Gamma(-\2-{\epsilon\02}) \0 (2\pi^\2)^\epsilon \Gamma(-\2)}
\int_0^\infty dk (k^2+m^2)^{-\2+{\epsilon\02}}
=-(4-N){8\pi m\01+\epsilon}I
\ee
where $I$ is the divergent integral introduced in (\ref{g2ren}).

The classical mass of a monopole is given by $M^{\rm cl}=4\pi m_0/g_0^2$.
In the renormalization scheme given by (\ref{g2ren}), 
where $m=m_0$ and $g_0^{-2}-g^{-2}=2(4-N)I$, we have the counterterm
\be\label{deltaM}
\delta M\equiv 4\pi m(g_0^{-2}-g^{-2})=8(4-N)\pi m I.
\ee

Adding (\ref{Ebulkres}) and (\ref{deltaM}) we find that
the divergent contributions from $I^{\rm div}=-{1\08\pi^2}{1\0\epsilon}$
cancel, but there is a finite remainder for $N=2$,
\be\label{Mcorr}
M^{(1)\rm bulk}+\delta M=-(4-N){m\0\pi}+O(\epsilon),
\ee
which has been overlooked in Refs.~\cite{Kaul:1984bp,Imbimbo:1985mt},
claiming vanishing quantum corrections.
While the renormalization conditions
of Refs.~\cite{Kaul:1984bp,Imbimbo:1985mt}
were identical to the ones we specified in sect.~\ref{sec:Z},
Ref.~\cite{Kaul:1984bp} did not specify the
regularization method used to obtain its null result,
and Ref.~\cite{Imbimbo:1985mt} regularized by inserting
slightly different
oscillatory factors in the two-point function and in the
integral over the spectral density in (\ref{Ebulkres}), 
a procedure that is not evidently self-consistent.

As we shall discuss in sect.~\ref{sec:central},
the finite remainder (\ref{Mcorr}) is associated with
an anomalous contribution to the $N=2$ central charge
of equal magnitude. The renormalization scheme set up
in sect.~\ref{sec:Z} is special in that in this scheme
all of the quantum correction to the monopole mass
are equal to the scheme-independent anomalous contribution
of the central charge, whereas for other schemes
one could also have further nonanomalous contributions
to both mass and central charge. However, before
we can conclude that (\ref{Mcorr}) is the final result for the quantum
corrections to the susy monopole mass, we have to
discuss also the surface contributions. 

\subsection{Surface contributions}
\label{sec:surfterms}

The surface terms for the susy monopole were obtained
in (\ref{T00surf});
they contain no contributions from fermions, only from
the vector, scalar, and ghost fields
\bea\label{M1surf}
M^{(1)\rm surf}&=&\int d^3x
T_{00}^{(2){\rm tot.deriv.}}=\int d^3x \biggl[
\4\6_j^2 \< a_0^2+a_{\3S}^2+2bc\> \nn
&&\qquad\qquad
-\2\6_j\6_k\<a_j a_k\>+2\6_j\<a_j \6_0 a_0\>\biggr],
\eea
where $j,k=1,2,3$. The difference between $N=2$ and $N=4$ is
only the range of the index $\3S\,$: $\3S=1,2,3,5,6$ for
$N=2$, and $\3S=1,2,3,5,\ldots,10$ for $N=4$.
We shall show that the sum of all surface contributions
cancels for the $N=2$ monopole, but for $N=4$ the extra four 
(pseudo-)scalars
yield a new type of divergence which we will have to dispose of.

To evaluate (\ref{M1surf}) we need the propagators for
the various fields in the monopole background.
As one can see from the linearized bosonic fluctuation equations
(\ref{boslineqam}) and (\ref{boslineqrest}),
the propagators for the fields $a_{\9m}=(a_m,s)$
are given by
\be\label{aaprop}
\<a^b_{\9m}(x)a^c_{\9n}(y)\>
=i\<x|[(-\6_0^2\delta^{bc}+(D_{\9k}^2)^{bc})\delta_{\9m\9n}
+2g\epsilon^{bac}F^a_{\9m\9n}]^{-1}|y\>
\ee
whereas the remaining (pseudo-)scalar fields ($p$ for $N=2$; $p$ and $q_I$
for $N=4$)
all have the same propagator
\be
\<p^b(x)p^c(y)\>=
i\<x|[-\6_0^2\delta^{bc}+(D_{\9k}^2)^{bc}]^{-1}|y\>, 
\ee
As explained below (\ref{E0pt}), for $a_0$ we have 
\be
\<a^b_0(x)a^c_0(y)\>=-\<p^b(x)p^c(y)\>,
\ee
and the anticommutative nature of the ghosts leads to
\be
\<b^b(x)c^c(y)\>=-\<c^c(y)b^b(x)\>=-\<p^b(x)p^c(y)\>.
\ee

The covariant derivatives $(D_{\9k}^2)^{ac}=(D_{k}^2)^{ac}+
(gS\times(gS\times))^{ac}$ have a complicated form when written out
\be
(D_{\9k}^2)^{ac}=\6_k^2\delta^{ac}+2\epsilon^{abc}gA_k^b\6_k
+g^2(A_j^aA_j^c-\delta^{ac}A_j^bA_j^b)+g^2(S^aS^c-\delta^{ac}S^2).
\ee
However, we only need the asymptotic values of the propagators.
Substituting the asymptotic values (\ref{asympt})
of the background fields we obtain
\bea
(D_{\9k}^2)^{ab}&\to&\6_k^2\delta^{ab}
+{2\0r}(\hat x^a \6^b-\hat x^b\6^a)
-{1\0r^2}(\delta^{ab}+\hat x^a \hat x^b)\nn
&&-m^2\left(1-{1\0mr}\right)^2(\delta^{ab}-\hat x^a \hat x^b)
\eea
where we recall that $m=gv$ and $\hat x^i\equiv x^i/r$.
The term $gF^a_{\9m\9n}$ appearing additionally in
(\ref{aaprop}) has the asymptotic behavior given in (\ref{asympt})
\bea
gF^a_{mn}\to -\epsilon_{mnk}{\hat x^a \hat x^k\0r^2},
\quad gF^a_{m6}=-
gF^a_{6m}=g(D_mS)^a\to -{\hat x^a x^m\0r^2}.
\eea

Because the
propagator $\<a_ja_0\>$ vanishes and
thus all nontrivial
surface terms in (\ref{M1surf}) involve two spatial derivatives,
transforming (\ref{M1surf}) into an integral over the sphere at infinity
we have
\be
\int d^3 x \6_i\6_j\ldots=\lim_{r\to\infty}
r^2\oint d\Omega \hat x_i \6_j \ldots,
\quad
\int d^3 x \6_j^2 \ldots=\lim_{r\to\infty}
r^2\oint d\Omega {\6\0\6r} \ldots
\ee
Hence, we need only keep the terms of order $1/r$ when
expanding the propagators around their common leading
term
\be
(-\6_0^2+D_{\9k}^2)^{-1ab}\to
{1\0\Box-m^2}(\delta^{ab}-\hat x^a \hat x^b)
+{1\0\Box}\hat x^a \hat x^b + O({1\0r}).
\ee
This straight away disposes of the terms with $F_{\9m\9n}$. Moreover,
we can also drop the terms proportional
to ${1\0r}(\hat x^a\6^b-\hat x^b \6^a)$ because the
derivatives $\6^a$ and $\6^b$ either act on a factor ${1\0r}$
and then produce terms falling off like $1/r^2$ or they
produce a term proportional to a single loop momentum 
in the numerator (see below) in which case symmetric integration
in momentum space gives a vanishing result.
Hence,
all propagators can be simplified to essentially the same form
\be\label{asbkprops}
\<a_M^a(x)a_N^b(y)\>
\to \eta_{MN}G^{ab}(x,y),\quad
\<b^a(x)c^b(y)\>\to -G^{ab}(x,y)
\ee
with
\be
G^{ab}(x,y)=
\<x|
\left[{-i\0-\Box+m^2-{2m\0r}}(\delta^{ab}-\hat x^a \hat x^b)
+{i\0\Box}\hat x^a \hat x^b \right]|y\>.
\ee

With these results at hand, we find for the contribution
of the surface terms to the mass of the $N=2$ monopole
\bea\label{M1surf2}
M^{(1)\rm surf,N=2}&=&\lim_{r\to\infty}{1\04}4\pi r^2
{\6\0\6r}\<a_0^2+a_j^2+s^2+p^2+2bc-2s^2\>\nn
&=&(-1+3+1+1-2-2)\lim_{r\to\infty}\pi r^2{\6\0\6r}\<s^2\>=0.
\eea
Thus in the $N=2$ case the contributions from the surface
terms in the quantum Hamiltonian cancel completely.

On the other hand, in the $N=4$ case $\<a_{\3S}^2\>$ in (\ref{M1surf}) involves
four extra scalar fields and there is no longer
a cancellation of these surfaces terms. We instead find
\be\label{M1surf4}
M^{(1)\rm surf,N=4}=4\times{1\04}\lim_{r\to\infty}4\pi r^2{\6\0\6r}\<s^2\>
\ee
so that we have to evaluate the expression on the right-hand side.
Up to an irrelevant constant we find in the limit of large $r$
\be
\<s^2\>=G^{aa}(y,x)|_{y=x}
\simeq 2\,\<y|{-i\0-\Box+m^2-{2m\0r}}|x\>|_{y=x}+{\rm const.},
\ee
where the factor 2 is due to the trace of $(\delta^{ab}-\hat x^a \hat x^b)$.
Following the procedure of section \ref{sec:indexth} to evaluate
such matrix elements by inserting complete sets of plane wave states we obtain
\bea
\<s^2\>&\simeq&2\int {d^{4+\epsilon}k\0(2\pi)^{4+\epsilon}}{-i\0(k^2+m^2)+2ik^\mu\6_\mu-\6_\mu^2
-{2m\0r}}\nn
&\simeq&2({2m\0r})\int {d^{4+\epsilon}k\0(2\pi)^{4+\epsilon}}{-i\0(k^2+m^2)^2}
=4{m\0r}I
\eea
with $I$ the logarithmically divergent integral defined
previously in (\ref{g2ren}).
Hence,
\be\label{M1surf4res}
M^{(1)\rm surf,N=4}=-16\pi m I.
\ee

In the $N=4$ theory there is no wave function or coupling constant
renormalization that could produce a counterterm from the
renormalization of the classical monopole mass. However, as we
have mentioned in sect.~\ref{sec:Z}, there is in general
the need for (additive) composite-operator
renormalization. The $N=2$ theory is special in that
the susy current multiplet is finite in the sense that
standard wave function and coupling constant renormalization suffices.
In the $N=4$ theory, however, the unimproved susy current
multiplet receives the counterterm (\ref{deltajunimpr})
involving improvement terms.
Since the expression for $T_{00}$ that we have derived in sect.~\ref{sec:bulksurf}
is unimproved, we have additive renormalization through
\be
\label{dt}
\delta T_{00}^{\rm comp.op.ren.}=-(Z_{\Delta T}-1)\Delta T_{00}^{\rm impr}
\ee
with $Z_{\Delta T}=Z_{\Delta U}$ and the latter given by (\ref{ZDeltaU}).
According to (\ref{Tmunuimpr}), the improvement term for the
energy density reads
\be\label{T00impr}
\Delta T_{00}^{\rm impr}=-{1\06}\6_j^2(A_{\mathcal J}A^{\mathcal J})
\ee
with $\mathcal J=5\ldots 10$. This gives
\bea
\delta M^{\rm comp.op.ren.}&=&-(Z_{\Delta U}-1)\int d^3x \left(-{1\06}\right)
\6_j^2(A_{\mathcal J}A^{\mathcal J})\nn&=&-(Z_{\Delta U}-1)\int d^3x \left(-{1\06}\right)
\6_j^2(S^2)\nn
&=&\lim_{r\to\infty}-(Z_{\Delta U}-1)\left(-{1\06}\right)4\pi r^2 {\6\0\6r}
v^2\left(1-{2\0mr}\right)\nn&=&-(12g^2I)\left(-{4\pi m\03g^2}\right)
=+16\pi mI,
\eea
which completely cancels (\ref{M1surf4res}).

As we have remarked in sect.~\ref{sec:Z},
the improved susy current multiplets do not require
composite-operator renormalization. Indeed,
had we included the surface terms (\ref{T00impr}) in (\ref{M1surf}),
we would have found complete cancellation in (\ref{M1surf4}),
because the improvement term involves 6 scalar fields in $N=4$
and an overall factor $-{1\06}$.

Note that in $N=2$ we had complete cancellation of the
surface term contributions already in the unimproved energy density.
In this model, and only there, the improvement
terms themselves are finite operators. With $\mathcal J$ running
only over two values, we have
\be
\int d^3x \< -{1\06}\6_j^2 a_{\mathcal J}^2 \>
=-2\times{1\06}\lim_{r\to\infty} 4\pi r^2 {\6\0\6r}\<s^2\>=
+{16\03}\pi m I
\ee
but this is now compensated by coupling constant and wave function
renormalization, since with the value
of $Z_S$, eq.~(\ref{ZS}), in the Feynman-$R_\xi$ gauge we also have
\be
(Z_S-1)\int d^3x \left(-{1\06}\right)
\6_j^2(S^2)=(4g^2I)\left(-{4\pi m\03g^2}\right)=-{16\03}\pi m I.
\ee

To summarize, we have found that the quantum corrections from the surface
terms in the Hamiltonian to the mass of the monopole
cancel completely 
upon composite-operator renormalization, which is
nontrivial in $N=4$.
The net result for the one-loop
quantum mass is then given by the bulk
contributions corresponding to the sum over zero-point energies
plus a standard counter\-term from the classical monopole mass,
eq.\ (\ref{Mcorr}),
which is nontrivial only in $N=2$,
\be
M^{(1)}_{N=2}=-{2m\0\pi};\qquad
M^{(1)}_{N=4}=0.
\ee

\section{Central charge}
\label{sec:central}
In the previous sections we obtained for the $N=2$ model a (negative) 
finite one-loop
correction to the monopole mass from the bulk (\ref{Mcorr}), whereas for the $N=4$ 
model we found divergent one-loop surface contributions. The latter 
were  cancelled by a composite operator renormalization which is proportional to the
improvement term for the energy momentum tensor.

The non-vanishing corrections raise the question whether and how 
they agree with
BPS saturation. Naively one would
expect that the central charge (\ref{UV},\ref{UUVV}) as a topological object
receives no quantum correction. 
The situation is however more complicated: both
in $N=2$ and in $N=4$ a direct one-loop calculation
for the central charge operator (\ref{UV},\ref{UUVV}) gives 
\cite{Imbimbo:1985mt,Rebhan:2004vn}
\begin{eqnarray}
  \label{u1}
	U^{(1)}&=&U^{(cl)}+
\2\int d^3 x\, \6_i \left(
S^a \epsilon_{ijk} \langle
F^a_{jk}[A+a]-F^a_{jk}[A] \rangle \right)\nonumber\\
&=&U^{(cl)}+\hal \int d^3x\,\epsilon_{ijk}\varepsilon_{abc}
	  \partial_i(g\bg{S^a}\langle a_j^b a_k^c)\rangle\nonumber\\
	       &=&\frac{4\pi m}{g_0^2}-16 \pi m I \ ,
\end{eqnarray}
where $I$ was given in (\ref{g2ren}). This result
is due to a $\langle a_i a_j\rangle$ loop, which is
evaluated using (\ref{aaprop}) and steps similar to the ones
leading to (\ref{M1surf4res}), but now it
is the same for the $N=2$ and $N=4$ model. 
According to (\ref{g2ren})
the coupling constant renormalizes
as $\frac{1}{g_0^2}-\frac{1}{g^2}=2(4-N)I$, 
so that in the $N=2$ case there is 
no one-loop correction, whereas in the $N=4$ model 
the UV-divergence in (\ref{u1}) remains uncanceled.\\

{\bf{$N=4$ monopole}}. In the $N=4$ model a detailed analysis showed that the
operator $U$ needs a composite operator renormalization which is proportional
to the improvement term (\ref{UimprN4}) \cite{Rebhan:2005yi}:
\begin{equation}
 \label{uren}
	\delta U^{\rm{comp.op.ren.}}= 4g^2I\biggl[U+
	   \frac{i}{8}\int d^3x\,\partial_i(\varepsilon^{ijk}
	    \bar\lambda\alpha^1\gamma_{jk}\lambda)\biggr].
\end{equation}
Taking this counterterm into account in (\ref{u1}) one obtains a vanishing
one-loop correction for the central charge of the $N=4$ monopole. 
This observation
resolved the puzzle of an apparent divergent one-loop correction to the
central charge \cite{Rebhan:2005yi}. Together with the 
results of section \ref{sec:surfterms} this restores 
not only the BPS saturation  $M=U$ at the one-loop order  but also the 
super multiplet structure
of the generators of the susy algebra. Both energy momentum tensor 
and central charge receive composite operator renormalizations proportional 
to improvement terms (\ref{dt}, \ref{uren}) through one-loop surface terms. These 
composite operator counterterms form a susy multiplet \cite{Rebhan:2005yi}.\\

{\bf{$N=2$ monopole}}. For the $N=2$ monopole the situation 
is more complicated. There is no way to balance the finite mass correction 
$\Delta M^{(1)}=-\frac{2m}{\pi}$ we obtained in (\ref{Mcorr}) by a one-loop contribution from 
(\ref{u1}). In \cite{Rebhan:2004vn} we have shown that the central 
charge despite its 
topological nature receives a one-loop correction which is associated with
an anomaly of the (conformal) central charge current. 

The anomaly in the (conformal) central charge turns out
to be  a member of the anomaly multiplet  besides the trace and the super 
conformal anomaly. In our approach, where the theory is dimensionally regularized
by embedding it in a higher dimensional space, this anomalous contribution
to the central charge appears as a non-vanishing momentum flow of fermions in 
the regulating extra dimension. Due to quantum effects the violation of
parity in the regulating five dimensional space has a finite reminder
as an anomalous central charge contribution in the four dimensional world.

According
to the choice (\ref{4dgamma}) the central charge $U$ stems from the 
$T^{05}$ component
of the six-dimensional ``energy momentum'' tensor, including the 
symmetric and anti-symmetric part of the r.h.s. of (\ref{djN2}) in 
this notion\footnote{Using $\Gamma_7\Gamma^{PQMRS}=\varepsilon^{PQMRST}\Gamma_T$
all terms of the r.h.s. of (\ref{djN2}) are proportional to a single gamma matrix.}.
The antisymmetric part gave the ordinary central charge $U$. From
the symmetric part, i.e. the genuine, momentum in the regulating extra dimension, 
one thus gets potentially non-vanishing contribution to the central 
charge (\ref{UV}). 

The mode functions and thus 
the propagators of the bosonic fields (\ref{eqn:bosfluct})
are even in the extra momentum and therefore give no contributions
to $T^{05}$. 
This is not the case for the fermionic fields and thus 
propagators (\ref{eq:n2psi}). Using the mode expansion (\ref{eq:fqf})
we obtain
\begin{eqnarray}
  \label{eq:an1}
  \langle T_{05}^{\mathrm{ferm}}\rangle&=&
  \langle \bar\psi\gamma_0\partial_5\psi\rangle\nonumber\\
  &=&-\int\frac{d^\epsilon\ell}{(2\pi)^\epsilon}\int\frac{d^3k}{(2\pi)^3}
   \frac{\ell^2}{2\omega}(|\chi^+_k|^2-|\chi_k^-|^2)(x)\ 
\end{eqnarray}
Here we have omitted
terms linear in the extra 
momentum $\ell$, as they vanish by symmetric integration,
and contributions from massless modes (zero modes),
which give scale-less integrals that do not contribute in 
dimensional regularization. 
Integration over space again leads to the difference
of spectral densities $\Delta\rho$ 
defined in (\ref{drh}) and evaluated in (\ref{drh2}).
The anomalous contribution to the central charge 
thus becomes 
\bea\label{eq:an2}
U_{\rm anom}&=&\int d^3x\, \langle \Theta_{05} \rangle =
\int {d^3k\,d^\epsilon\ell \0 (2\pi)^{3+\epsilon}}
{\ell^2 \0 2 \sqrt{k^2+\ell^2+m^2}}\,\Delta\rho(k^2) \nonumber\\
&=&-4m
\int_0^\infty {dk\02\pi} 
\int {d^\epsilon\ell \0 (2\pi)^{\epsilon}}
{\ell^2 \0 (k^2+m^2)\sqrt{k^2+\ell^2+m^2}}\nonumber\\
&=&-8 {\Gamma(1-{\epsilon\02})\0(4\pi)^{1+{\epsilon\02}}}
{m^{1+\epsilon}\01+\epsilon}
=-{2m\0\pi}+O(\epsilon).
\eea
The correction (\ref{eq:an2}) 
matches exactly the mass correction (\ref{Mcorr}) so that the BPS bound is saturated at the one-loop level,
but in a very nontrivial way.

The nonzero result (\ref{eq:an2}) is in fact in complete accordance
with the low-energy effective action for $N=2$ super-Yang-Mills theory
as obtained by Seiberg and Witten \cite{Seiberg:1994rs,Seiberg:1994aj,Alvarez-Gaume:1997mv}.\footnote{We are grateful
to Horatiu Nastase for pointing this out to us.}
According to the latter, the low-energy effective action is fully determined
by a prepotential $\mathcal F(A)$, which to one-loop order is
given by
\be\label{F1loop}
\mathcal F_{\rm 1-loop}(A)={i\02\pi}A^2\ln {A^2\0\Lambda^2},
\ee
where $A$ is a chiral superfield
and $\Lambda$ the scale parameter of the theory generated by
dimensional transmutation. 
The value of its scalar component $a$
corresponds in our notation to $gv=m$. In the absence of a $\theta$
parameter, the one-loop renormalized coupling is given by
\be\label{taua}
{4\pi i\0g^2}=\tau(a)={\partial^2 \mathcal F \0 \partial a^2}
={i\0\pi}\left(\ln{a^2\0\Lambda^2}+3\right).
\ee
This definition agrees with the on-shell renormalization scheme
that we have considered above, because the latter involves only the
zero-momentum limit of the two-point function of the massless fields.
For a single magnetic monopole, the central charge 
is given by
\be
|U|=|a_D|=\left|{\partial \mathcal F\0\partial a}\right|=
{1\0\pi}a\left(\ln{a^2\0\Lambda^2}+1\right)
={4\pi a\0g^2}-{2a\0\pi},
\ee
and since $a=m$, this exactly agrees with the result
of our direct calculation in (\ref{eq:an2}).

The low-energy effective action associated with (\ref{F1loop})
has been derived from a consistency requirement with the
anomaly of the U(1)$_R$ symmetry of the microscopic theory,
which forms a multiplet with the trace anomaly and a new anomaly
in the conformal central charge, which is responsible for
the nonzero correction (\ref{eq:an2}).

\section{Conclusions}

In this article we have calculated the one-loop corrections
to the mass and central charge of $N=2$ and $N=4$ susy monopoles in
3+1 dimensions. Besides the nonvanishing anomalous contribution
in the $N=2$ case missed in the older literature,
the calculation in the $N=4$ case involved two novel features:
surface terms contributing to the mass corrections and
composite operator renormalization of both the mass and the
central charge. With everything taken into account, we explicitly
verified BPS saturation at the quantum level.

\TABLE{%
\begin{tabular}{c||c|c|c|c|c||c}
~ & \multicolumn{2}{c|}{bulk~~~~~~} & surf. & ordinary & comp.op. & total \\[-3pt]
~ & nonan. & anomalous & terms & c.t. & renorm. & result \\
\hline
\hline
\rule[-3pt]{0pt}{16pt}
$N=2$, $M^{(1)}$ & $xI$ & $-{2m/\pi}$ & 0 & $-xI$ & 0 & $-{2m/\pi}$ \\
\rule[-3pt]{0pt}{16pt}
$N=2$, $U^{(1)}$ & 0 & $-{2m/\pi}$ & $xI$ & $-xI$ & 0 & $-{2m/\pi}$ \\
\hline
\rule[-3pt]{0pt}{16pt}
$N=4$, $M^{(1)}$ & 0 & 0 & $xI$ & 0 & $-xI$ & 0 \\
\rule[-3pt]{0pt}{16pt}
$N=4$, $U^{(1)}$ & 0 & 0 & $xI$ & 0 & $-xI$ & 0 \\
\hline
\hline
\end{tabular}
\caption{Individual contributions to the one-loop corrections
to mass ($M^{(1)}$) and central charge ($U^{(1)}$) of
$N=2$ and $N=4$ monopoles. Here 
$x=-16\pi m$ and $I$ is the divergent integral defined
in (\ref{g2ren}).}
}

Table 1 summarizes our findings by listing the individual contributions
to the one-loop corrections
to mass and central charge of the
$N=2$ and $N=4$ monopoles.
The mass contributions involve ``bulk'' terms which can be identified
as the familiar sums over zero-point energies. These cancel in
$N=4$, but in $N=2$ give a divergent contribution containing
a finite anomalous contribution that is left over after standard
coupling constant renormalization
(denoted by ``ordinary c.t.''). As we have discussed above,
in our scheme of dimensional regularization this anomalous
contribution appears
as a remainder of parity violation in the fifth dimension,
which is in turn caused by the fact that the fermionic mode functions
$\chi^+_k$ and $\chi^-_k$ in (\ref{eq:fqf}) come with normalization
factors $\sqrt{\omega+\ell}$ and $\sqrt{\omega-\ell}$, respectively,
where $\ell$ is the momentum of modes along the fifth dimension.
This asymmetry is made possible by the existence of two inequivalent
representations of the Dirac algebra in 5 dimensions and leads
to a net momentum flow $\<T_{05}\>$ which in 3+1 dimensions
becomes the anomalous contribution to the central charge.
Just as the trace anomaly in $T_\mu{}^\mu$ is the anomaly
in the conservation of the conformal stress tensor $x^\nu T_{\nu\mu}$,
and $\gamma\cdot j_{\rm susy}$ is the anomaly in the conformal susy
current $\7x j_{\rm susy}$, also $U_{\rm anom}$ is the anomaly
in the conformal central charge current, in perfect analogy
to the situation in the susy kink \cite{Rebhan:2002yw}.

The nonanomalous part of the central charge is given by the
usual surface term in 3+1 dimensions. Its one-loop correction
is divergent and is cancelled the counterterm induced by standard
coupling constant renormalization in the case of $N=2$, but
in the finite $N=4$ theory, it is cancelled by infinite
composite operator renormalization as we have shown in detail
in Ref.~\cite{Rebhan:2005yi}.
In the present paper we have shown that there are also surface terms
in the mass formula, due to partial integration of the
bosonic terms of the form $\6\varphi\6\varphi$ in the
gravitational stress tensor, while the usual sum over zero-point
energies corresponds to bulk terms of the form $-\varphi\6^2\varphi$.
The surface terms $\6(\varphi\6\varphi)$ 
give divergences that precisely match the composite operator
renormalization of the susy current multiplet.

The need for composite operator renormalization may be a little
surprising. However, nonrenormalization theorems for conserved
currents are restricted to internal symmetries (like the
vector current in the CVC theory), and do not hold in general
for space-time symmetries. The energy-momentum tensor generally
requires nonmultiplicative renormalization
through improvement terms 
\cite{Callan:1970ze,Freedman:1974gs,Freedman:1974ze,Collins:1976vm,Brown:1980pq}. Curiously, in the $N=2$ theory both the unimproved and the
improved stress tensor is finite (i.e., does not require
composite operator renormalization), whereas in the ``finite''
$N=4$ theory only the improved stress tensor is finite.
The standard (textbook) result for the classical monopole mass refers
to the unimproved stress tensor, which is also what is obtained
by dimensional reduction from higher dimensions.
In 3+1 dimensions, one could also start with an improved
stress tensor, and this would also make
the $N=4$ theory free from composite operator
renormalization of the susy current multiplet. 
Note that this reduces the classical value of the
mass of a monopole to 2/3 of its standard value \cite{Rebhan:2005yi}.

It would be interesting to study the complete structure of the
multiplets of the local improved and nonimproved currents,
both in $x$-space and in superspace. In this connection it may be
relevant to note that the stress tensor one obtains in the susy
multiplet of currents differs from the gravitational stress tensor
by a total derivative.

\acknowledgments

We would like to thank Robert Sch\"ofbeck for collaboration
in the early stages of this work. We also thank M.\ Bianchi, S.\ Kovacs,
and Ya.\ Stanev for discussions. We gratefully acknowledge
financial support from the C.\ N.\ Yang Institute
for Theoretical Physics at SUNY Stony Brook, 
the Technical University of Vienna,
and NSF grant no. PHY-0354776.

\providecommand{\href}[2]{#2}\begingroup\raggedright\endgroup
\end{document}